\documentclass[final,2p,12pt,sort&compress]{elsarticle}
\usepackage[T1]{fontenc}
\usepackage[utf8]{inputenc}

\usepackage{graphicx}
\usepackage{xcolor}

\newcommand{\unit}[1]{\,\mathrm{#1}}

\usepackage{amssymb,amsmath}
\DeclareMathOperator{\sgn}{sgn}

\usepackage{xurl}
\journal{Biomedical Signal Processing and Control}

\usepackage{mathptmx}
\usepackage{etoolbox}
\usepackage{multirow}
\usepackage{makecell}
\usepackage{graphicx}
\usepackage{dcolumn}
\usepackage{bm}
\usepackage{xr-hyper}
\usepackage[%
colorlinks = true,
urlcolor   = blue,
linkcolor  = black
]{hyperref}
\usepackage{setspace}
\usepackage[left=2cm,right=2cm,top=2cm,bottom=2.5cm]{geometry}
 
\usepackage{pgffor}
\usepackage{xassoccnt}

\begin{document}

\begin{frontmatter}

\title{Multifractal organization of EEG signals\\ in Multiple Sclerosis}


\author[1,2]{Marcin W\k{a}torek}
\author[2,
]{Wojciech Tomczyk}
\fntext[fn1]{Present address (Wojciech Tomczyk): Institute of Materials Engineering, Faculty of Science and Technology, University of Silesia, 75 Pu\l{}ku Piechoty 1a, Chorz\'{o}w 41-500, Poland}
\author[3]{Magda Gaw\l{}owska}
\author[3]{Natalia Golonka-Afek}
\author[3]{Aleksandra \.{Z}yrkowska}
\author[4,5]{Monika Marona}
\author[4,5]{Marcin Wnuk}
\author[4,5]{Agnieszka S\l{}owik}
\author[2,6]{Jeremi K. Ochab}
\author[3]{Magdalena Fafrowicz}
\author[7]{Tadeusz Marek}
\author[8,2,6]{Pawe\l{} O\'{s}wi\k{e}cimka\corref{cor1}}
\ead{pawel.oswiecimka@ifj.edu.pl}
\cortext[cor1]{Corresponding author}

\affiliation[1]{
organization={Faculty of Computer Science and Telecommunications, Cracow University of Technology},
addressline={Warszawska 24}, 
city={Kraków},
postcode={31-155},
country={Poland}}
            
\affiliation[2]{
organization={Institute of Theoretical Physics, Jagiellonian University},
addressline={\L{}ojasiewicza 11},
city={Kraków},
postcode={30-348},
country={Poland}}

\affiliation[3]{
organization={Department of Cognitive Neuroscience and Neuroergonomics, Jagiellonian University},
addressline={\L{}ojasiewicza 4},
city={Kraków},
postcode={30-348},
country={Poland}}

\affiliation[4]{
organization={Department of Neurology, Jagiellonian University Medical College},
addressline={Jakubowskiego 2},
city={Kraków},
postcode={30-688},
country={Poland}}

\affiliation[5]{
organization={Department of Neurology, University Hospital in Krakow},
addressline={Jakubowskiego 2},
city={Kraków},
postcode={30-688},
country={Poland}}

\affiliation[6]{
organization={Mark Kac Centre for Complex Systems Research, Jagiellonian University},
addressline={\L{}ojasiewicza 11},
city={Kraków},
postcode={30-348},
country={Poland}}

\affiliation[7]{
organization={Faculty of Psychology, SWPS University},
addressline={Techników 9},
city={Katowice},
postcode={40-326},
country={Poland}}

\affiliation[8]{
organization={Complex Systems Theory Department, Institute of Nuclear Physics, Polish Academy of Sciences},
addressline={Radzikowskiego 152},
city={Kraków},
postcode={31-342},
country={Poland}}

\begin{abstract}
Quantifying the complex/multifractal organization of the brain signals is crucial to fully understanding the brain processes and structure. In this contribution, we performed the multifractal analysis of the electroencephalographic (EEG) data obtained from a controlled multiple sclerosis (MS) study, focusing on the correlation between the degree of multifractality, disease duration, and disability level. Our results reveal a significant correspondence between the complexity of the time series and multiple sclerosis development, quantified respectively by scaling exponents and the Expanded Disability Status Scale (EDSS). Namely, for some brain regions, a well-developed multifractality and little persistence of the time series were identified in patients with a high level of disability, whereas the control group and patients with low EDSS were characterised by persistence and monofractality of the signals. The analysis of the cross-correlations between EEG signals supported these results, with the most significant differences identified for patients with EDSS $> 1$ and the combined group of patients with EDSS $\leq 1$ and controls. No association between the multifractality and disease duration was observed, indicating that the multifractal organisation of the data is a hallmark of developing the disease. The observed complexity/multifractality of EEG signals is hypothetically a result of neuronal compensation -- i.e., of optimizing neural processes in the presence of structural brain degeneration.
The presented study is highly relevant due to the multifractal formalism used to quantify complexity and due to scarce resting-state EEG evidence for cortical reorganization associated with compensation. 
\end{abstract}



\begin{keyword}
multifractal\sep time series\sep EEG\sep nonlinearity\sep complexity
\end{keyword}

\end{frontmatter}


\section*{Introduction}

Multifractality is a concept that has been successfully applied across many disciplines, ranging from social sciences to exact sciences \cite{Stanley1988,Ivanov1999,LOPES2009634,IHLEN2013633,DROZDZ201632,Shao2016,Jiang_2019,Watorek2021,Augustyniak2022,harte2001multifractals}. The physical observables recorded in these systems often form complex time series exhibiting nontrivial statistical properties. In particular, the singular time series is analyzed from the point of view of singularity strength. From the methodological perspective, the basic assumption is to employ a set of scaling exponents to describe the stochastic processes and, thus, singularity related to the multiscaling properties of the data. The standard characteristic obtained from this methodology is the multifractal spectrum where singularity strength is quantified by the exponent $\alpha$ and $f(\alpha)$, which refers to the singularity strength and Hausdorff dimension of the subset support with specific $\alpha$~\cite{halsey1986}. This concept is particularly useful when nonlinear dependencies, not quantified by commonly used methods such as autocorrelation function or Fourier transform, are present in the signal. The nonlinearity of time series corresponding to the level of data complexity can be quantified by the width of the multifractal spectrum. The wider the spectrum, the more complex the data structure. In recent years, several algorithms have been proposed to estimate the multifractal spectrum in the empirical data~\cite{Kantelhardt2011}. Among them, two main approaches can be distinguished. One is based on the wavelets methodology, where singular behavior can be detected by employing the wavelet transform and quantifying its scaling properties among different scale ranges~\cite{muzy1993,wendt2007}. The alternative one uses scaling properties of the variance estimated on different scale ranges, which connects it directly to the diffusion theory \cite{KANTELHARDT200287}. Both these approaches are robust and commonly used; however, in the authors' experience, the variance-based methodology of estimating the multifractal spectrum exhibits more stability, whereas wavelets are more useful when local scaling properties of the signals are investigated~\cite{oswiecimka2006}. 

Physiological systems are commonly considered complex systems exhibiting nonlinear, multiscale organization. Multifractal formalism becomes particularly useful in this area, especially when singular measures or functions, common in experimental data, are considered~\cite{Eke2002,West2010,PittmanPolletta2013,Dutta2014,catrambone2021}. One example in this respect is heart rate dynamics. The fluctuations of healthy subjects' cardiac interbeat time series reveal nonlinear, multifractal behavior that is difficult to distinguish from the less complex pathological state on the grounds of classical methodology~\cite{Goldberger2002}. The identified multifractal complexity indicates the heart rate control mechanism is similar to physical cascade-like processes such as turbulence. Such findings motivate a new, multiscale approach to quantify and model physiological systems under healthy and pathological conditions. The multifractal methodology has also been applied to analyze human gait rhythm in the case of neurodegenerative disease~\cite{CHATTERJEE2020123154}. The multifractal cross-correlation analysis has been used to quantify the correlation between the stride-interval fluctuations between the left and right foot. It has shown that the multiscale behavior characterizes healthy and diseased patients; however, the degree of multifractality is higher in the case of healthy individuals compared to patients with amyotrophic lateral sclerosis (ALS) disease. One of the flagship examples is the analysis of electroencephalography (EEG) data collected from the human brain \cite{doi:10.1073/pnas.1007841107,LOPES2009634,Racz2019,10.3389/fphys.2018.01767,10.1371/journal.pone.0068360}. Significant efforts have been made in recent years to resolve many issues, using a multifractal framework, regarding the interpretation of EEG data from patients struggling with different neurological disorders, for example, Alzheimer \cite{JEONG20041490,SCHWILDEN200631,Zorick2020}, epileptic seizures \cite{Schelter2006,Ghosh2014,Zhang2015,David2020}, schizophrenia \cite{Slezin2007,Dick2022} or sleep apnea \cite{Lee2002,Ma2018}. Especially in the long run, it seems that multifractal analyses are good candidates for solid prognostic and diagnostic tools in biomedical signal processing \cite{10.3389/fphys.2018.01767,Eke2002,Dutta2014,CHATTERJEE2020123154,Racz2020,Valentim2021,Lavanga2021}. Multifractal Detrended Fluctuation analysis is one of those. There are no such studies, however, concerning multiple sclerosis (MS). This work constitutes a major step toward this unexplored subject.

The correlation structure of the signals can also be quantified with regard to the time scale and amplitudes of the fluctuations separately~\cite{oswiecimka_detrended_2014}. This methodology is beneficial when the analyzed signal is not fractal, however, it can also be used to support and amend the results of the multifractal analysis. Moreover, it can be regarded as an enhancement of the standard analysis of  EEG time series employing the Pearson correlation coefficient, where the temporal signal organization is quantified without signal decomposition.

Multiple sclerosis is a chronic immune-mediated disease, the most common non-traumatic disorder of the central nervous system (CNS) \cite{goldenberg2012multiple,browne2014atlas,dobson2019multiple}. The pathological hallmark of MS is demyelination and subsequent axonal degeneration that results in CNS lesions \cite{bitsch2000acute,lucchinetti2000heterogeneity}. These neural alterations are present even in patients with early-stage MS \cite{lucchinetti2011inflammatory}. Symptoms depend on the lesion areas, therefore, the common presenting ones include fatigue, optic neuritis, depression, heat sensitivity, dizziness, numbness, loss of balance, and cognitive dysfunction \cite{krupp1988fatigue,calabresi2004diagnosis,chiaravalloti2008cognitive,dobson2019multiple,KARACA2022231}. Cognitive impairment in MS has prevalence rates of 34 to 70\% \cite{chiaravalloti2008cognitive,benedict2020cognitive}. The cognitive decline is reported in measures of processing speed and episodic memory, complex and sustained attention, information processing efficiency, executive functioning, verbal fluency, conceptual reasoning, and visuospatial perception \cite{rao1991cognitive,chiaravalloti2008cognitive,benedict2020cognitive}.
Due to the anatomical nature of disease-induced changes, one of the primary tools used for both MS diagnosis and further monitoring is magnetic resonance imaging (MRI) \cite{bakshi2008mri,filippi2010contribution,sahraian2010role,wattjes20212021}.
EEG can be used as a complementary method to study secondary disease-induced changes in MS, such as cognitive impairment and other functional declines. Previous research on MS using EEG suggests that the method allows studying functional changes of the brain not only in voluntary movement \cite{leocani2001fatigue} or cognitive tasks \cite{whelan2010impaired,keune2017exploring,torabi2017diagnosis}, but also in a resting-state condition \cite{gschwind2016fluctuations,vecchio2017electroencephalography}. The alterations are observable even in the early stages of the disease~\cite{zipser2018cortical}. 

\section*{Materials and methods}

\begin{figure}[htp!]
\includegraphics[width=\textwidth]{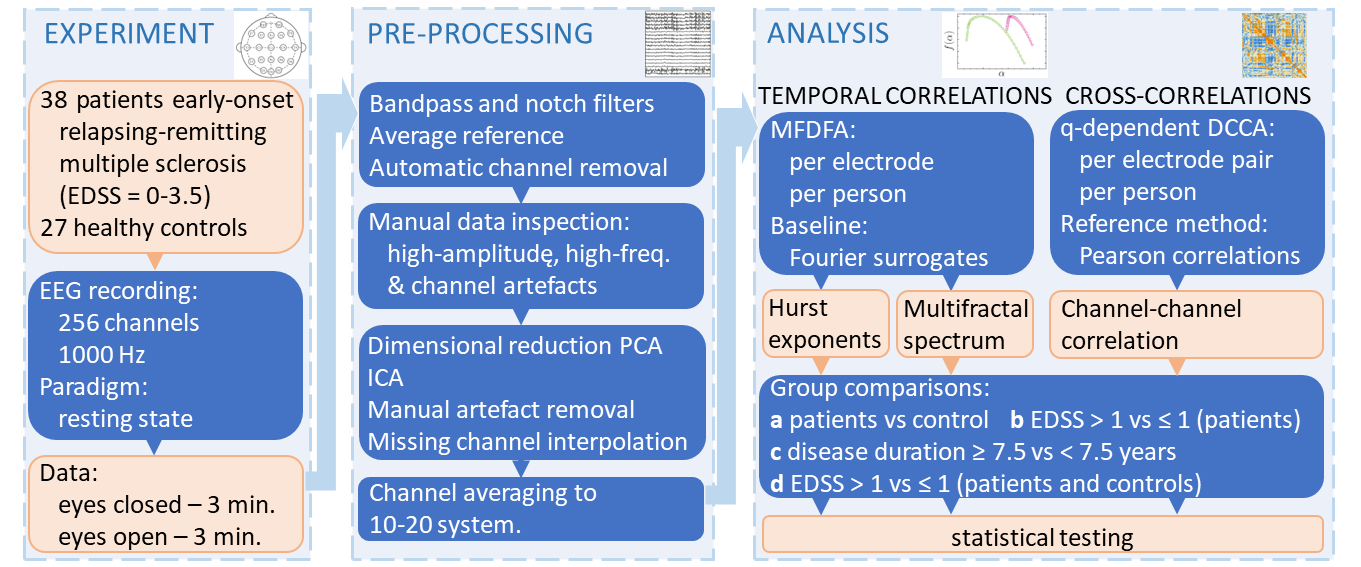}
\caption{Flowchart summarizing data collection, processing and analysis.}
\label{fig_flowchart}
\end{figure}

For the overall summary of data collection, processing and analysis, see Fig.~\ref{fig_flowchart}. Each stage is described in detail in the following subsections.

\subsection*{Participants and experimental design}
The presented analyses were performed on a group of 38 multiple sclerosis patients (mean age: 34.3 ${\pm}$ 2.97, 19 females) 
and 27 healthy controls (mean age: 35.6 ${\pm}$ 2.79, 16 females). 
The patients were recruited by attending physicians from the cohort of patients from the Jagiellonian University Multiple Sclerosis Clinic, whereas healthy controls were recruited using online advertisements. Prior to participating in the study, all patients were diagnosed with early-onset relapsing-remitting multiple sclerosis (RRMS) with Expanded Disability Status Scale (EDSS) scores~\cite{Kurtzke1444} ranging from 0 to 3.5 points (mean: 1.2 ${\pm}$ 0.84). 
The participants were right-handed, with normal or corrected-to-normal vision, and without a history of substance dependence. The study protocol was approved by the Research Ethics Committee at the Institute of Applied Psychology of the Jagiellonian University and was carried out in accordance with the Helsinki Declaration.
Each subject gave their written informed consent. EEG was recorded in eyes open and eyes closed conditions -- each lasting 3 minutes. Participants were instructed to remain as still as possible during the recording.
For the eyes open condition, subjects were asked to focus on a cross presented on the computer screen, whereas in eyes closed they were asked to close their eyes and relax. The experiment was designed and presented using E-Prime 2.0$^\text{\textcopyright}$ (Psychology Software Tools).

\subsection*{EEG recording and preprocessing}
Continuous dense-array EEG data (HydroCel Geodesic Sensor Net, EGI System 300; Electrical Geodesic Inc., OR, USA) were collected using 256-channel EEG (sampled at 1000 Hz, bandpass filtered at 0.01 to 100 Hz with a vertex electrode as a reference) and recorded with NetStation Software (Version 4.5.1, Electrical Geodesic Inc., OR, USA). The impedance for all electrodes was kept below 50 k$\Omega$. Off-line data analysis was performed with the open-source EEGLAB toolbox (http://sccn.ucsd.edu/eeglab). Data were digitally filtered to remove frequencies below 0.5 Hz and a notch filter was applied to remove the 50 Hz frequency, i.e., the mains frequency. The average reference was recomputed and bad channels were automatically removed by kurtosis measures with a threshold value of 5 standard deviations. Continuous data were then visually inspected to manually remove the remaining bad channels or time epochs containing high-amplitude, high-frequency muscle noise, and other irregular artefacts.
Independent component analysis was used to remove artefacts from the data. Due to a large number of channels, the decomposition of the EEG data with the Infomax algorithm was preceded by Principal Component Analysis. Fifty independent components were extracted and visually inspected for each subject. Based on the spatiotemporal pattern, components recognised as blinks, heart rate, saccades, muscle artefacts, or bad channels were removed. Missing channels were interpolated, and ICA weights were recomputed.

\subsection*{Multifractal Detrended Fluctuation Analysis}
Multifractal Detrended Fluctuation Analysis~\cite{KANTELHARDT200287} is a generalisation of the Detrended Fluctuation Analysis\cite{PhysRevE.49.1685}. It has been successfully deployed as a robust tool that facilitates the multilevel
characterisation of time series (see examples of its use~\cite{oswiecimka2006, Oswiecimka2020,Ochab2022}). The basics of MFDFA can be briefly summarised in the following steps. First, for a particular time series $\{x(i)\}_{i=1}^{N}$ on a compact support, the integrated signal profile $X(j)$ is calculated according to the formula:
\begin{equation}\label{eq:1}
    X(j)=\sum_{i=1}^{j}(x(i)-\langle x\rangle),\quad j=1,\dots,N,
\end{equation}
where $\langle\dots\rangle$ stands for averaging over the entire time series.

From then on, $X(j)$ is divided into $M_s$ nonoverlapping segments $\nu$ of length $s\,(s<N)$ starting from both ends of the time series (ergo into $2M_s$ such segments in total). 
For each segment $\nu$, the local trend can be approximated by fitting an $m$-th order polynomial $\left(P^{(m)}_\nu\right)$ and subtracted from the data ($m$ governs the effectiveness of the method~\cite{Oswiecimka2013}). Subsequently, the detrended variances for all segments $\nu$ and the respective segment lengths $s$ can be computed:
 \begin{equation}
     F^2(\nu,s)=\frac{1}{s}\sum_{j=1}^s\left\{X((\nu-1)s+j)-P^{(m)}_\nu(j)\right\}^2
 \end{equation}
 for segments $\nu=1,\dots,M_{s}$ and
 \begin{equation}
     F^2(\nu,s)=\frac{1}{s}\sum_{j=1}^s\left\{X(N-(\nu-M_s)s+j)-P^{(m)}_\nu(j)\right\}^2
 \end{equation}
for segments $\nu=M_s+1,\dots,2M_s$.

Ultimately, $F^2(\nu,s)$ is averaged over $\nu$s and the $q$-th order fluctuation function is calculated for all possible segment lengths:
\begin{equation}
    F_q(s)=\left(\frac{1}{2M_s}\sum_{\nu=1}^{2M_s}\left[F^2(\nu,s)\right]^{q/2}\right)^{1/q},\quad q\in\mathbb{R}\backslash\{0\}.
    \label{eq:fluct_function}
\end{equation}
The pivotal feature of $F(s)$ is the manifestation of power-law-type behavior (over a wide range of $s$, as in Fig.~\ref{fig_1}\textbf{a}) for a signal with fractal properties:
\begin{equation}
    F_q(s)\sim s^{h(q)}.
\end{equation}
As an outcome of the MFDFA procedure, one gets a family of exponents $h(q)$, the so-called generalised Hurst exponents~\cite{Mandelbrot1968}, which for a multifractal signal form a decreasing function of $q$ in opposition to a monofractal signal, where $h(2)\equiv H=\textrm{const.}$ For short-range correlated time series $H\sim 0.5$, whereas for long-range monofractal-correlated time series $H$ deviates from 0.5 and two intervals can be distinguished: $0<H<0.5$ (antipersistent signal) and $0.5<H<1$ (persistent signal).

Furthermore, based on the generalised Hurst exponents, one can obtain the multifractal/singularity spectrum of the H\"{o}lder exponents, i.e. $f(\alpha)$, by the following relations \cite{halsey1986}:
\begin{equation}
    \alpha=h(q)+qh^\prime(q),\quad f(\alpha)=q[\alpha-h(q)]+1,
    \label{eq:spectrum}
\end{equation}
where $h^\prime$ denotes a derivative of $h$, $\alpha$ determines the strength of the singularities, and $f(\alpha)$ can be viewed as the fractal dimension of a subset of the time series with singularities of magnitude $\alpha$. Furthermore, for positively correlated signals, the spectrum is shifted toward $\alpha>0.5$, and vice versa, the spectrum located below $\alpha<0.5$ indicates negative data autocorrelation. The maximum spectrum at $\alpha=0.5$ suggests weak linear correlations or lack thereof.

\subsection*{Fourier Surrogates}
Fourier surrogates are commonly used to test the statistical significance of the results of the multifractal analysis. This procedure is based on the comparative analysis of artificial data that are generated by randomly shuffling the phases of the Fourier transform of the original signal while preserving its amplitudes~\cite{schreiber2000} and performing the inverse Fourier transform. When this procedure is applied, the potential nonlinear correlations responsible for true multifractality in the signal are removed, but the linear ones are preserved. Therefore, only the monofractal behavior of the time series (narrow multifractal spectrum) should be recovered from the surrogate data.

\subsection*{$q$-Dependent Detrended Cross-Correlation Coefficient}
In recent years, MFDFA has been generalised to the case of two time series and further to Multifractal Cross-Correlation Analysis (MFCCA)\cite{oswiecimka_detrended_2014}, and subsequently, Detrended Cross-Correlation Analysis~\cite{PhysRevE.92.052815} has been derived with the cross-correlation coefficient denoted as $\rho(q,s)$. This coefficient describes the cross-correlations between a pair of time series $x(i)$ and $y(i)$, both on the particular time scale $s$ and regarding the amplitude of fluctuations filtered by $q$. Given two time series $X(j)$ and $Y(j)$ (cf.~Eq. (\ref{eq:1})), the covariance takes the form
\begin{equation}    
F_{xy}^2(\nu,s)=\frac{1}{s}\sum_{k=1}^s\left(X((\nu-1)s+k)-P^{(m)}_{X,\nu}(k)\right)\times\left(Y((\nu-1)s+k)-P^{(m)}_{Y,\nu}(k)\right)
\end{equation}
for $\nu=1,\dots,M_{s}$ and
\begin{equation}
F_{xy}^2(\nu,s)=\frac{1}{s}\sum_{k=1}^s\left(X(N-(\nu-M_s)s+k)-P^{(m)}_{X,\nu}(k)\right)\times\left(Y(N-(\nu-M_s)s+k)-P^{(m)}_{Y,\nu}(k)\right)
\end{equation}
for $\nu=M_s+1,\dots,2M_s$.

The $q$-th order covariance function reads
\begin{equation}
    F^q_{xy}(s)=\frac{1}{2M_s}\sum_{\nu}^{2M_s}\sgn\left(F_{xy}^2(\nu,s)\right)\left|F_{xy}^2(\nu,s)\right|^{q/2}.
\end{equation}
The coefficient $\rho(q,s)$ can be calculated accordingly\cite{ZEBENDE2011,PhysRevE.84.066118}
\begin{equation}
    \rho(q,s)=\frac{F^q_{xy}(s)}{\sqrt{F^q_{xx}(s)F^q_{yy}(s)}},\quad\rho(q,s)\in[-1,1]\;\textrm{for}\;q>0,
    \label{rho}
\end{equation}
where $F^q_{xy}(s)$ denotes the fluctuation function of the detrended covariance of a time series pair: $x$ and $y$. $F^q_{xx}(s)$ and $F^q_{yy}(s)$ represent the detrended fluctuation function of $x$ and $y$, respectively. Eq.~(\ref{rho}) can be regarded as the generalisation of the Pearson coefficient~\cite{Pearson1895,Rodgers1988} sensitive to correlations with respect to the signal amplitude and the scale considered.

\section*{Results}

\subsection*{Multifractal analysis}
  
We performed a multifractal analysis to characterise the nonlinear temporal organisation of the time series recorded by the EEG electrodes. 
To estimate multifractal characteristics, we applied a well-established method: multifractal detrended fluctuation analysis (MFDFA). Since the time series coming from adjacent electrodes in the high-density recording are strongly correlated, to reduce the complexity of calculations and the impact of noise on the analysis results, we collected the electrodes into 20 groups according to the international 10 -- 20 system (electrode group positions are shown in Supplementary Fig.~S1 online). 
Thus, the results obtained for the electrodes in the high-density recording were averaged within each of the 20 locations to which they belonged.

\begin{figure}[htp!]
\centering
\includegraphics[width=\textwidth]{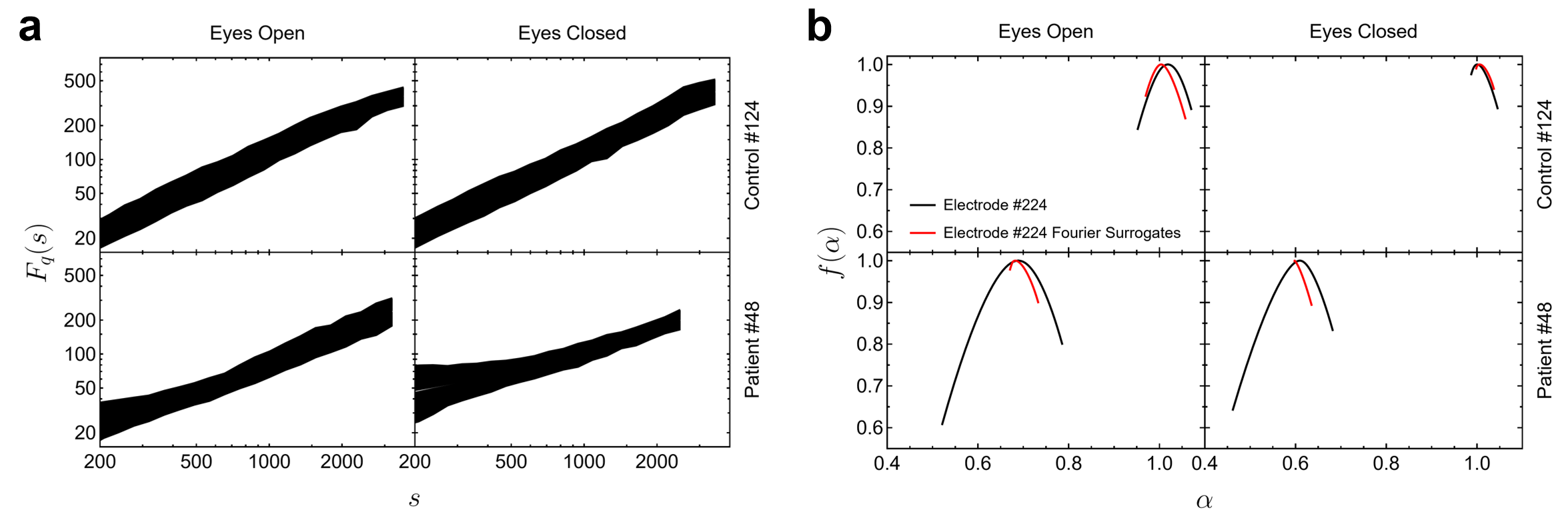}
\caption{\textbf{a} Fluctuation functions $F_q(s)$ with a visible power-law dependence. \textbf{b} Multifractal spectra $f(\alpha)$ estimated from fluctuation functions of the original time series (black curves) and its Fourier surrogates (red curves) in the scale range $s=200-2000$. The examples come from electrode \#224 (F4) of a single patient (\#48) and a control group participant (\#124). The patient's spectral widths are significantly wider than those of the surrogates, indicating the multifractality of the signal. The location of the control participant's spectrum $\alpha \gg 0.5$ characterises the signal as persistent.}
\label{fig_1}
\end{figure}

\begin{figure}[htp!]
\includegraphics[width=\textwidth]{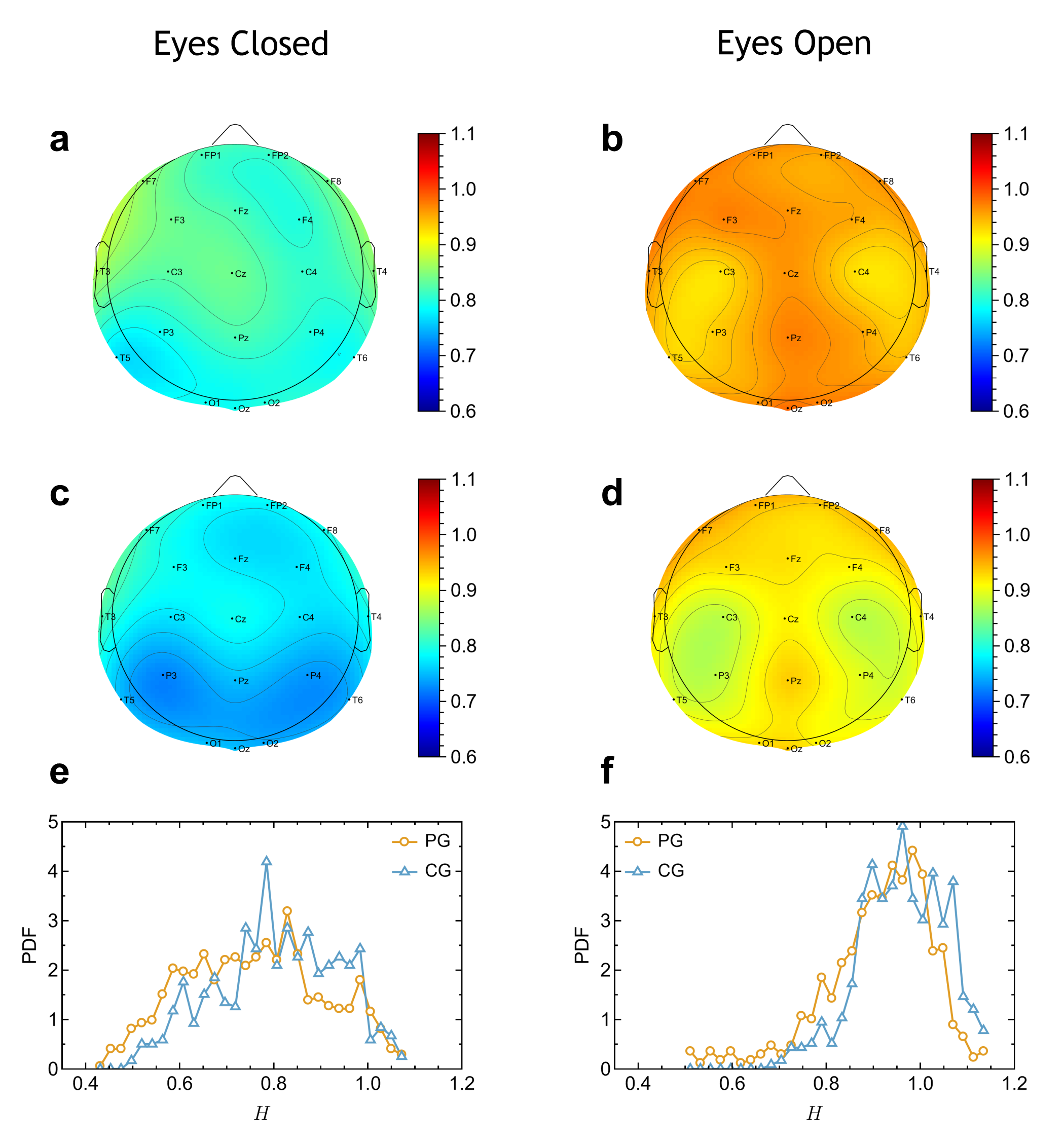}
\caption{The topographic distributions show group-averaged values for (\textbf{a}-\textbf{b}) controls and (\textbf{c}-\textbf{d}) patients. (\textbf{e}-\textbf{f}) The histograms show probability density functions (PDFs) of Hurst exponents collected from the twenty electrodes and all subjects within a group of patients, PG, and controls, CG.}
\label{fig_2}
\end{figure}

\begin{figure}[htp!]
\includegraphics[width=\textwidth]{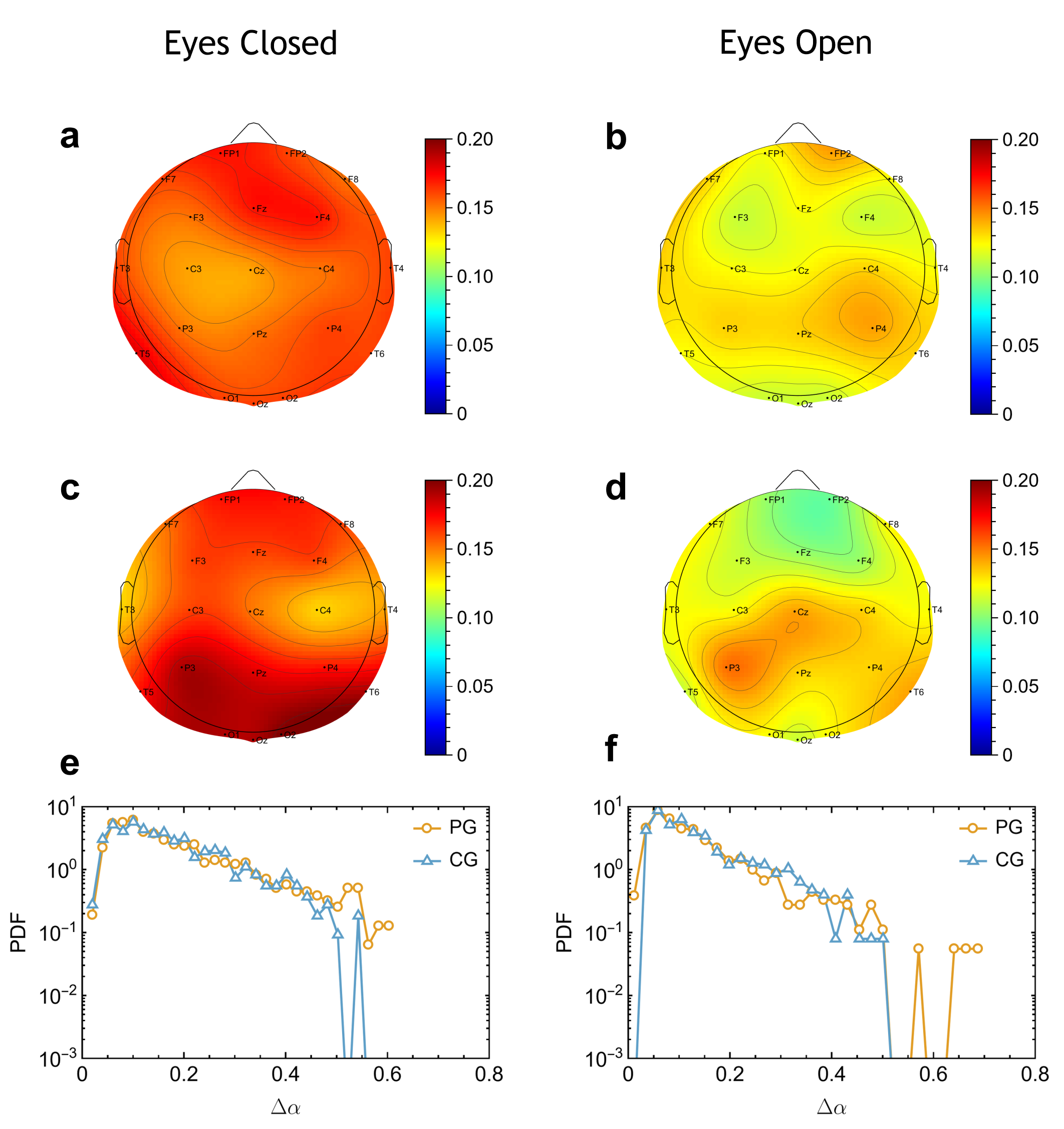}
\caption{The topographic distributions show group-averaged values for (\textbf{a}-\textbf{b}) controls and (\textbf{c}-\textbf{d}) patients. (\textbf{e}-\textbf{f}) The histograms show probability density functions (PDFs) of spectral widths collected from the twenty electrodes and all subjects within a group of patients, PG, and controls, CG.}
\label{fig_3}
\end{figure}

Results for sample electrodes and subjects, i.e., fluctuation functions $F_q(s)$ and multifractal spectra $f(\alpha)$ (see, Eq.~\ref{eq:fluct_function} and \ref{eq:spectrum}) for open and closed eyes, are shown in Fig.~\ref{fig_1}, panels a and b, respectively. In the considered scale range, the family of the fluctuation functions $F_q(s)$ clearly reveals power-law behavior for all considered $q$, which confirms the fractal organisation of the data. The estimated multifractal spectra for the presented functions $F_q(s)$ take the shape of an asymmetrical parabola and are located at $\alpha \gg 0.5$, indicating the strong persistence of the time series. Furthermore, the estimated spectra widths for the example patient assume $\Delta \alpha \approx 0.3$, indicating the multifractal organisation of the data. 
Comparison with the Fourier surrogates of the data, whose spectra are very narrow, confirms that the spectrum width of the original data is the effect of the nonlinear temporal data correlations and reflects the complex organisation of the time series. 
Group comparisons revealed several intriguing results. In Fig.~\ref{fig_2}, we show the topographical plots of average Hurst exponents estimated for the controls and patients at each electrode group considered.
There is a clear difference between the results obtained for subjects with closed and open eyes. In the former case, the average Hurst exponent assumes values $H\approx0.8$, whereas, in the latter case, the exponents are closer to one, indicating $1/f$ dynamics of the data. Thus, the time series for eyes open reveal stronger persistence than the ones for eyes closed.
Furthermore, the Hurst exponents for the control group are higher compared to the patients. Quantitatively, this effect is shown in Figs.~\ref{fig_2}\textbf{e}-\textbf{f}, where the probability density function (PDF) of the set of exponents estimated for each subject within the group is depicted. The difference between the control group and the patients is most visible for closed eyes. A similar conclusion can be drawn when considering the width of the singularity spectrum (Fig.~\ref{fig_3}). The time series for eyes closed differs significantly from the signals collected from subjects with eyes open. In the former case, the average $\Delta \alpha \approx 0.2$ indicates multifractal dynamics of the data, whereas in the latter, time series are monofractals. Thus, the temporal organisation of the signals for subjects with closed eyes is more complex than the data collected from subjects with open eyes. When comparing the results between the control group and the patients, the broader spectra are the attribute of the patients, which suggests richer dynamics of the data in the latter case.  

\begin{figure}[htp!]
\centering
\includegraphics[width=\textwidth]{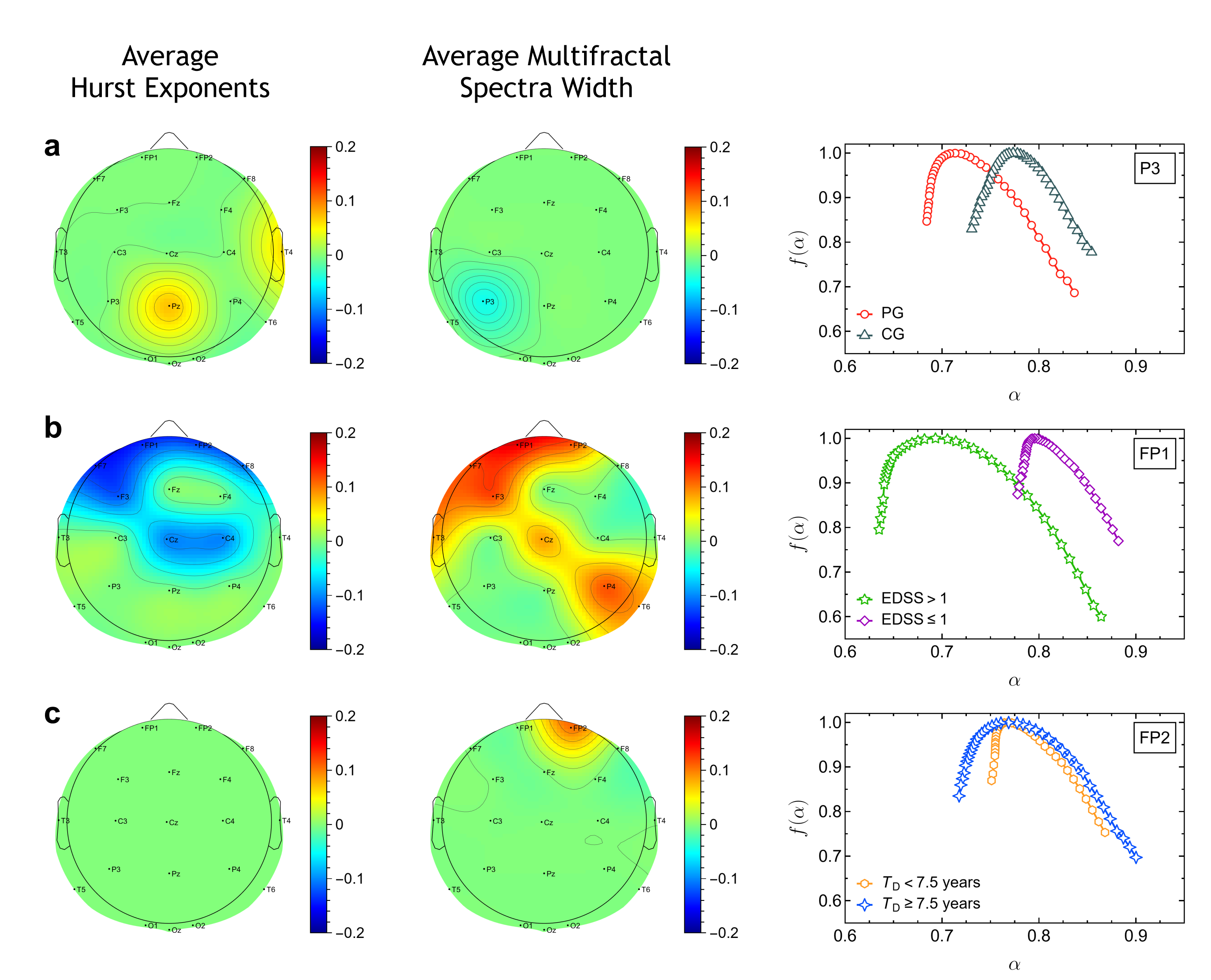}
\caption{Group differences between the Hurst exponents $H$ (left column) and multifractal spectra width $\Delta\alpha$ (middle column) taken over 20 electrodes for various cases:
\textbf{a} the control group and patients,
\textbf{b} patients with $\textrm{EDSS}>1$ and patients with $\textrm{EDSS}\le1$,
\textbf{c} patients with a time of disease longer than 7.5 years and shorter than 7.5 years. Right column: the average of multifractal spectra estimated for electrode P3, FP1, and FP2 for \textbf{a}, \textbf{b}, and \textbf{c} row, respectively. Only statistically significant results (see Supplementary Table~S2 for $p$-values) for eyes closed (see Supplementary Fig.~S2 for open eyes) are shown.} 
\label{fig_4}
\end{figure}

\begin{figure}[ht]
\centering
\includegraphics[width=.8\textwidth]{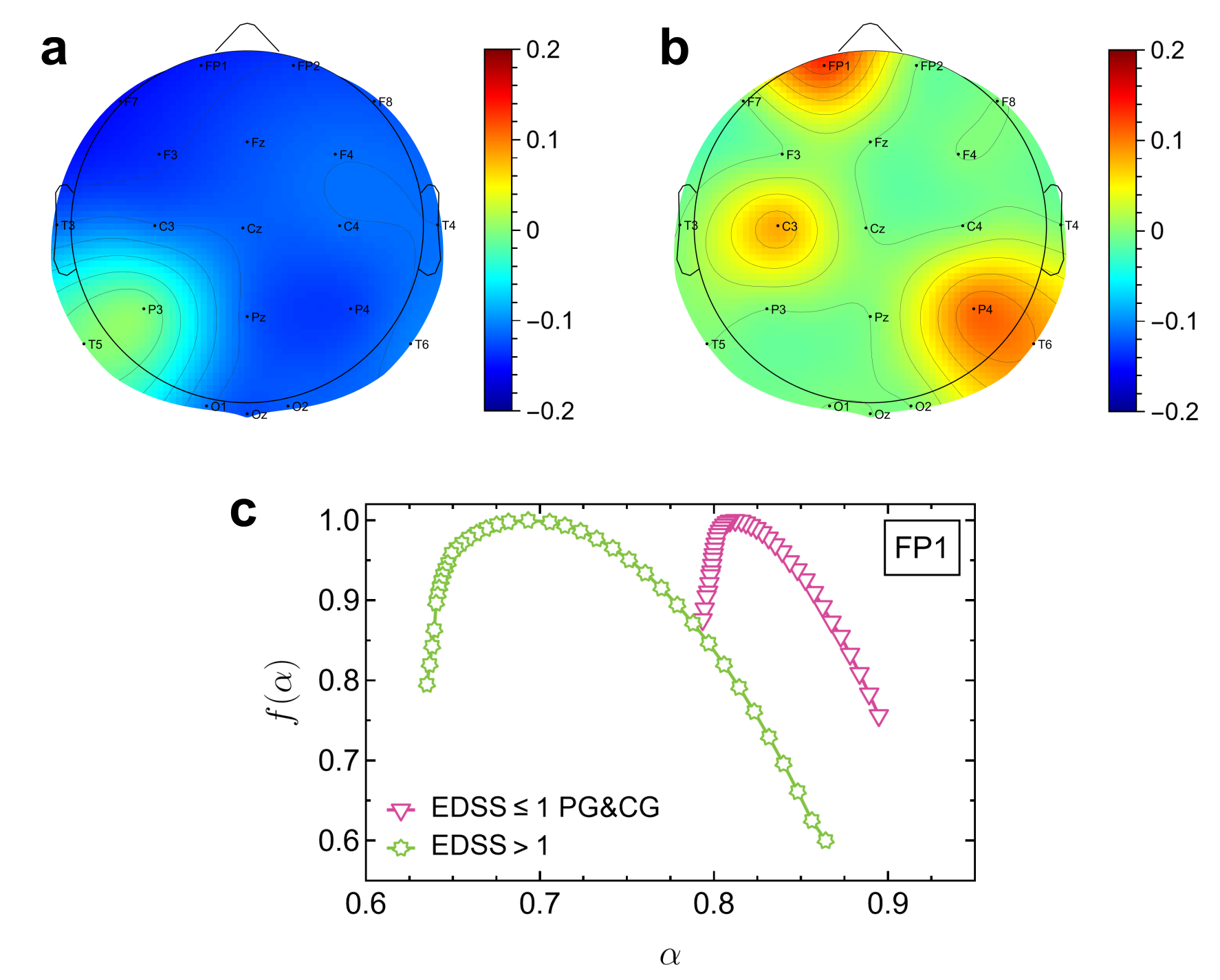}
\caption{The topographical plots show group differences between:
\textbf{a} group-average Hurst exponents and
\textbf{b} group-average multifractal spectra width $\Delta\alpha$.
\textbf{c} Group-average multifractal spectra estimated at electrode Fp1.
Only statistically significant results (see Supplementary Table~S2 for $p$-values) for eyes closed (see Supplementary Fig.~S2 for open eyes) are shown.} 
\label{fig_5}
\end{figure}

To highlight the differences between the groups, in Fig.~\ref{fig_4} we compare Hurst exponents $H$ and the multifractal spectrum width $\Delta \alpha$ directly between specific groups: (a) control and patient groups, (b) according to the Expanded Disability Status Scale (EDSS) \cite{Kurtzke1444}, patients with $\textrm{EDSS}>1$ and patients with $\textrm{EDSS}\le1$ and (c) patients with a disease time longer and shorter than 7.5 years. The figure shows the statistically significant differences in the average Hurst exponents (left side) and the average multifractal spectra (middle) between the control group and patients in the aforementioned cases. The $p$-values of Welch's $t$-tests for the results presented in Fig.~\ref{fig_3}-\ref{fig_4} are reported in Supplementary Tables S2-S3. Since the group comparisons (a)-(c) were pre-planned and only (d) was exploratory, we report $p$-values uncorrected for multiple comparisons. The highest differences between the groups are observed in case (b). For patients with $\textrm{EDSS}>1$, the values of $H$ are lower and, at the same time, the values of $\Delta \alpha$ are higher than for patients with $\textrm{EDSS}\le1$. Thus, the signals recorded for patients with higher $\textrm{EDSS}$ exhibit both less persistent signals but a more complex data organisation than the time series coming from patients with lower $\textrm{EDSS}$. In other cases, the differences are less visible, although in case (a) the exponents $H$ are also significantly higher for the control group compared to the patient group. In case (c), for only one electrode (FP2) $\Delta \alpha$ is higher for patients with a disease time longer than 7.5 years than for those with a shorter disease time. The estimated average spectra $f(\alpha)$ for the electrodes with the apparent difference between the groups considered are shown in Fig.~\ref{fig_4} (right column).
As indicated above, the most evident contrast between multifractal spectra has been identified for patients with different levels of disability EDSS (Fig.~\ref{fig_4}\textbf{b}). The time series for patients with higher disabilities are characterised by a more developed complexity of the signal and its lower persistence which is quantified by the width of the multifractal spectrum and its location, respectively. Motivated by this observation, we performed additional analysis and divided the results with respect to the level of disability, that is, one group consisted of the control group and patients with a lower level of disability $(\textrm{EDSS}\leq 1)$ and the other one of patients with a more developed disability $(\textrm{EDSS}>1)$.
The results of the statistically significant differences in the multifractal characteristics between these groups and the average multifractal spectra for the one electrode with high deviation are depicted in Fig.~\ref{fig_5}. In this case, the differences between the groups are even more evident. The group with none and lower disability is characterised by higher linear correlations (persistence) of the time series recorded from almost all electrodes. The exception is the electrodes T5 and P3, for which the difference is below the confidence level. Nonlinear analysis indicates that the electrodes P4, FP1 and C3 are crucial in distinguishing the groups. Signals at these electrodes exhibit a more developed hierarchical organisation reflected in the width of the multifractal spectra as presented in Fig.~\ref{fig_5}.

\subsection*{Cross-correlation analysis}
\label{ccanalysis}

To assess the coupling between electrode signals, we used both the Pearson coefficient and the detrended cross-correlation coefficient $\rho(q,s)$~(Eq.\ref{rho}). Since, in the latter case, our results were the most statistically significant, here we present the results for the coefficient $\rho$ with order $q=1$ and scale $s=200\unit{ms}$ that corresponds to frequency $5\,\unit{Hz}$. The results obtained for the Pearson coefficient and $\rho$ with $q=1$ and $s=400\unit{ms}$ are presented in Supplementary Figs.~S3 and S4 online. The exponent $q=1$  was chosen as the most meaningful, indicating the strongest correlation for medium-sized fluctuations. Similarly to the previous section, the results from the electrodes were grouped and averaged over the electrodes according to the 10 -- 20 system as shown in Supplementary Fig.~S1 online. In Fig.~\ref{fig_6}, the correlation matrices with the coefficient $\rho$ are depicted for the control group (Figs.~\ref{fig_6}\textbf{a}-\textbf{b}) and patients (Figs.~\ref{fig_6}\textbf{c}-\textbf{d}). The calculations were performed separately for the closed and open eyes conditions in the resting state.  

It is clear from Fig.~\ref{fig_6} that electrode signals are strongly correlated for both the control group and the patient group. The strong positive correlations characterise the signals from electrodes inside the brain areas considered, i.e., within the cerebral cortex's frontal, pre-frontal and occipital areas, and negative correlations are typical for signals coming from distinct regions. In Figs.~\ref{fig_6}\textbf{e}-\textbf{f}, we depicted the statistical distribution of the off-diagonal correlation coefficients of the control and patient group (left, L, and right, R, distribution tails separately). The clear difference between the left tail (negative cross-correlations) and the right one (positive cross-correlations) of the PDF can be noticed. The right tail of the distribution, both for the control group and the patients, has a similar shape with a maximum at $\approx 0.95$. For the left tail of the distribution, the maximum assumes smaller values and a small difference between the location of the maximum for the control group (0.9) and patients (0.83) can be noticed. 
To assess the average coupling between the electrodes, we also estimated the eigenvalue spectrum, and the collected results for all subjects are depicted in Figs.~\ref{fig_6}\textbf{g}-\textbf{h}. In most cases, the two distinct largest eigenvalues are observed for each individual, which form the two bulks clearly identified in Figs.~\ref{fig_6}\textbf{g} and \textbf{h}. These indicate the vital global component in the data shared by the brain signal. The difference between the control group and the patients is not noticeable.    

To highlight the differences in the results between the patients and the control group in Fig.~\ref{fig_7}, we filter the correlation matrices according to the statistical Welch's $t$-tests (all $p$-values are reported in Supplementary Fig. S6-S7). Namely, we leave the correlation coefficients only for the electrodes for which the average correlation difference between the considered groups is statistically significant at the confidence level $0.05$. 
To account for the multiple comparisons we estimate the false discovery rates (FDR) of these tests via $q$-values \cite{Storey2003} (see, Supplementary Table~S4), which however might be conservative considering the strong dependence of the cross-correlation values.
Similarly, as in the previous section, we compared the patients and the control group, as well as the groups with respect to the level of disability measured by EDSS and the duration of the disease. The results are visualized as a network of electrodes, where links denote the statistically significant differences between the considered groups at the given pair of electrodes. Moreover, the brain regions with the overall strongest cross-correlations are depicted in the right column of Fig.~\ref{fig_7}. As can be seen in row \textbf{b}, the most apparent difference can be observed in patients with high and low EDSS. 
The local false discovery rates (the Bayesian posteriors that the result is false positive) for the significant electrodes range between 0.57-0.84 in \textbf{b} and 0.30-0.66 in \textbf{d}.
Based on Fig.~\ref{fig_7}\textbf{b}, we identified the region of electrode P4 as the most significant in distinguishing between patients with different levels of EDSS.
In the other comparisons for closed eyes, the groups are barely distinguishable (FDR close to 1).
For open eyes (see Supp. Fig.~S5), the local FDRs range between 0.20-0.48 in \textbf{c} and 0.72-0.78 in \textbf{d}, and are close to 1 otherwise.
Pearson estimator of cross-correlation coefficients produces analogous, although less pronounced results, as presented in Supplementary Fig.~S8-S11.

\begin{figure}[htp!]
\centering
\includegraphics[width=.93\textwidth]{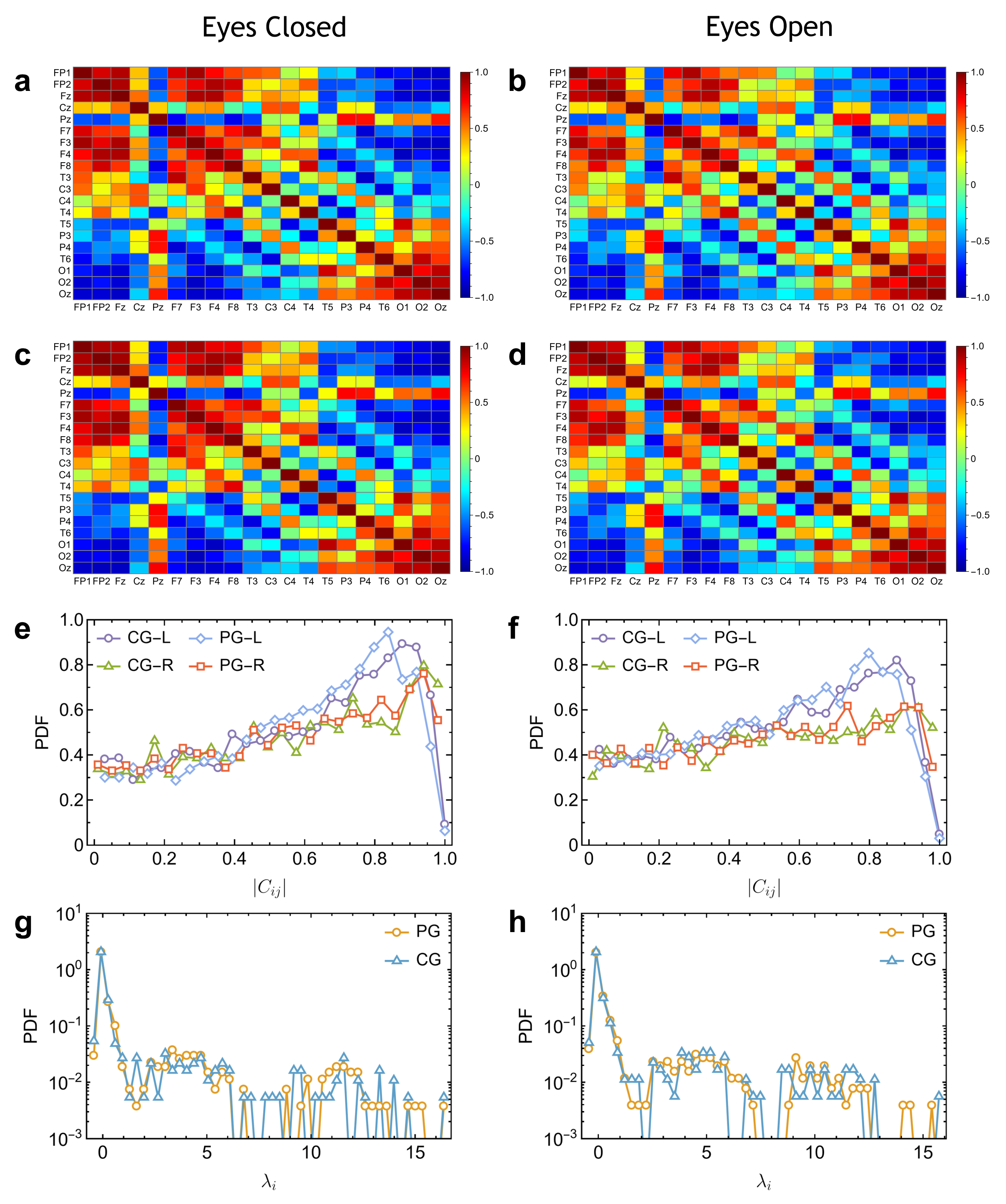}
\caption{Group-average detrended correlation matrices $\rho(q=1,s=200\,\textrm{ms})$ are shown for: \textbf{a-b} the control group (CG) and \textbf{c-d} patients (PG).
\textbf{e-f} Corresponding distributions of the off-diagonal matrix elements $C_{ij}$, with L -- left ($C_{ij}<0$), and R -- right tail ($C_{ij}\geq 0$) of the probability density function (PDF), respectively.
\textbf{g-h} Correlation matrix eigenvalues, $\lambda_i$ $(i=1,\ldots, 20)$, estimated for the patient and control groups.
The results are divided according to the experimental conditions: (\textbf{Left column}) eyes closed and (\textbf{Right column}) eyes open.
For Pearson correlations and DCC at another scale $s$, see Supplementary Figs.~S3-S4 online.} 
\label{fig_6}
\end{figure}
    
\begin{figure}[htp!]
\centering
\includegraphics[width=.7\textwidth]{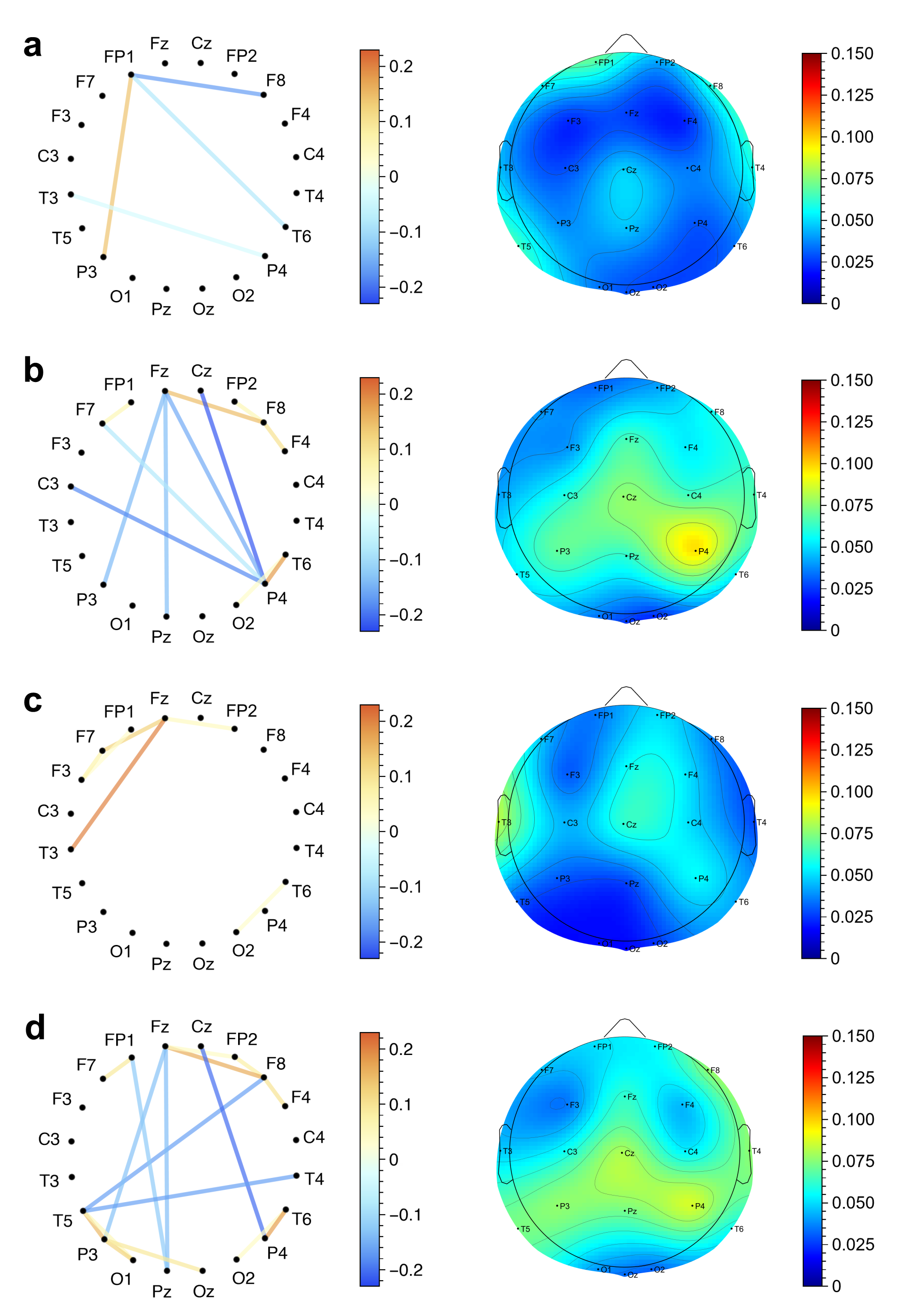}
\caption{(\textbf{Left column}) The links in the graph represent group differences in detrended cross-correlations $\rho(q=1,s=200\,\textrm{ms})$ between pairs of electrodes. The link's colour indicates the value of the difference. Only statistically significant links are shown, $p<0.05$.
(\textbf{Right column}) The topographic plots show the absolute differences at each electrode averaged over cross-correlations with all other electrodes. The group comparisons include:
\textbf{a} the control group and patients,
\textbf{b} patients with $\textrm{EDSS}>1$ and patients with $\textrm{EDSS}\le1$,
\textbf{c} patients with the disease duration $\geq$ 7.5 and $<7.5$ years, 
\textbf{d} patients with $\textrm{EDSS}>1$ and the combined group of patients with $\textrm{EDSS}\le1$ and controls.
The results are shown for the closed eyes condition. For results in open eyes condition, see Supplementary Fig.~S5 online).} 
\label{fig_7}
\end{figure}

\newpage
\section*{Discussion}
Multiple sclerosis is a chronic immune-mediated disease, characterised by demyelination and subsequent axonal degeneration resulting in CNS lesions. The current state of medical knowledge makes it possible to control symptoms and slow the progression of the disease, however, early recognition of symptoms and implementation of an appropriate treatment protocol becomes crucial for the effective management of MS. Interestingly, studies show that some changes in the brain activity observed in MS patients may be stage-specific. Faivre \textit{et al.}~\cite{Faivre2016} point at the initial increase in network connectivity. Networks become more flexible, which may serve as a compensatory mechanism, ensuring system efficacy in the early stages of the disease. The brain learns how to compensate for disease-related deficits by overwriting the existing neuronal connections~\cite{Fried2022}. Thus, often patients diagnosed with the early stages of MS do not present significant disruptions of cognitive functions. As MS progresses, brain networks become less adaptable and more prone to overload by the upcoming input. This shift from flexible to rigid results in the decline of cognitive functioning~\cite{SCHOONHEIM2022}. One of the methods used to quantify the disability in multiple sclerosis and monitor the severity of neurological changes over time is the 10--point Expanded Disability Status Scale (EDSS). 

In our paper, we performed a multifractal analysis of time series from patients diagnosed with multiple sclerosis and compared the results with those of the control group.
The analysed signals reveal a high order of temporal organisation, which a set of scaling exponents could quantify. Thus, the time series can be considered as (multi)fractals quantitatively described by the multifractal spectrum. We show that there is a statistically significant difference in signal complexity, measured by the width of the multifractal spectrum, between patients scoring less than or equal to 1 point on the EDSS scale –- thus not presenting any neurological symptoms -- compared to patients scoring above 1 to 3.5 points, with higher signal complexity in the latter. Thus, in the former case, the data are nonlinear and hierarchically organised, revealing a more prosperous multifractal organisation than in the latter case, with much poorer nonlinearity but the strongest linear dependencies quantified by Hurst exponent. Analysis of cross-correlations between electrode signals confirmed the difference in data organisation between patients with $\text{EDSS}>1$ and those without neurological symptoms, which is demonstrated on the level of significance of region coupling.    

Moreover, the analysis of the widths of the multifractal spectra revealed differences in activity within the recording sites reflecting the activity of the central executive (CEN) and the default mode network (DMN). Activity within the CEN has been associated with performance in tasks that require cognitive control and attention, whereas activity within the DMN has been associated with self-referential processing, mind wandering, and other forms of internally focused cognition. 
Although altered activity and connectivity within DMN and CEN have previously been reported in MS patients, the results are mixed. For example, a decrease in functional connectivity has been observed between the posterior cingulate cortex (PCC), considered the main hub of DMN, and other more frontal regions, as well as a decrease in functional connectivity within the CEN~\cite{Bonavita2011}. When comparing cognitively impaired MS patients to MS patients with preserved cognitive functions, studies report both increased~\cite{Meijer2018} and decreased~\cite{Eijlers2019, Rocca2018} functional connectivity within DMN. 
Considering the early strengthening of the connectivity hypothesis, our result may be interpreted as an early attempt of the system to compensate for the disease-related deficits.

The presented study comes with potential limitations and challenges. The first limitation relates to some aspects of the multifractal methodology used to assess the scaling exponents. The parameter $q$ used to calculate the fluctuation function must be chosen with particular care since the higher moments of $F_q(s)$ could diverge when the pdf of the times series is characterized by fat tails. In our study, we restricted the range of the $q$ parameter to fulfil this restriction. Moreover, the proper estimation of the Hölder exponent requires eliminating the possible trend in the data. Within the MFDFA algorithm and procedure of $\rho(q,s)$ coefficient estimation, detrending is performed by subtracting the fitted polynomial of the assumed degree. However, choosing the polynomial degree is a delicate matter since it can influence the results of correlation structure significantly and must be carefully considered. Based on the research and our experience, the most reasonable choice of polynomial order is 2 or 3~\cite{Oswiecimka2013}. 
The multifractal spectrum is a global measure of time series organization and provides average information about the temporal dependencies and structure of the data. Therefore, it is an excellent measure of the overall correlation skeleton. However, future studies could also consider the local scaling properties, i.e. localization of the singularity within the time series. This analysis could be performed through wavelet methodology and uncover additional complex properties of the data.

The second possible limitation of our study is its sample size. Our study was performed on a group of 38 multiple sclerosis patients and 27 healthy controls, which is a rather large group comparable with other investigations~\cite{kiiski2018,jamoussi2023}. We expect that extending the sample could make our conclusions even more evident. Moreover, considering patients with more significant Expanded Disability Status Scale scores could give us information about the change in the correlation characteristics with disease development. Therefore, our plans include developing a study analyzing dynamical changes in the complexity measures depending on the disease progression. 

Another possible limiting factor regards not considering the interaction of the physiological signals and EEG. There is an increasing awareness in the field of EEG research about the substantial impact of physiological signals, such as heart rate, on EEG data (e.g., ~\cite{RUIZPADIAL2018,catrambone2021,Lehnertz2021,Pernet2020,Mahmoodi2023}). Regrettably, our study was not designed to delve deeper into the interaction between these factors, and specifically, we did not independently record heart rate. Consequently, we lack the relevant data to incorporate this aspect into our analysis. In an effort to mitigate the influence of physiology on the presented results, we employed techniques like ICA decomposition to remove components associated with breathing, heart rate, etc. It is crucial to acknowledge that in future studies investigating signal complexity in MS patients due consideration should be given to accounting for the influence of physiological factors.

It is worth noting that, according to the literature, the complexity of the brain signal can be understood as not only the amount of noise mixed in the observed brain signal, which, if optimal, allows effective information processing~\cite{Faisal2008} but also the level of integration and segregation of brain networks and the ability to transition from one network to another~\cite{Cnudde2023}.

Cognitive impairment in MS patients is the effect of demyelination and axonal damage, which strongly influence the effectiveness of the conduction of neural impulses and lower brain activity in areas affected by pathological changes~\cite{10.1093/brain/awx022,lennea2013}. However, in the early stages of the disease development, despite the loss of structural neural integration, MS patients' cognitive functioning is on the ordinary level, undistinguished from the period before the disease began. Thus, it is assumed that the observed stability of cognitive functioning is the effect of the neuronal mechanisms called compensation, which is responsible for optimizing the neural processes in the presence of structural neuronal degeneration. Although the process is still poorly recognized due to conceptual and experimental obstacles, we can assume the hypothesis that compensation is reflected in brain activity and thus is potentially identifiable with the EEG technique. The observed complex behavior of the signal of MS patients quantitatively characterized by multifractal characteristics could be the effect of the abovementioned mechanism. The brain reorganizing to maintain functioning at the appropriate level, e.g., taking over the function of the defective brain regions by other ones, must increase the signal's complexity. The multiscale hierarchical structure of the brain tissues is yet another facet influencing the multifractal organization of the brain signal. Thus, we suggest that an increase in the multifractality of MS patients' signals reflects the processes of compensation occurring in the brain. If we combine data from previous studies, pointing to compensatory mechanisms in early-onset MS, and our results, showing a higher complexity of the EEG signal in patients who function relatively well despite the diagnosis, the higher complexity can be interpreted in terms of adaptive information processing mechanisms and the brain's ability to learn how to deal with emerging lesion-related deficits on the neuronal level.  

Moreover, our study revealed the vital role of the signals from the right hemisphere (recorded from electrode P4), distinguishing between patients in the group with EDSS $\leq$ 1 and those with EDSS > 1. In this context, the results from Lenne and colleagues, who quantified cortical communication in the EEG resting state of both cognitively deficient and non-deficient patients with RRMS, are noteworthy. The study demonstrates that mutual information in the right hemisphere serves as an indicator of compensatory processes~\cite{lennea2013}. While studies on neural compensation using resting-state data are relatively rare, an asymmetric right hemisphere pattern of compensation in neurodegeneration has also been observed in studies related to Huntington's and Alzheimer's diseases. Gregory and colleagues, using both task and resting-state fMRI data of premanifest Huntington disease patients and employing a novel cross-sectional model of compensation, indicated the role of right hemisphere activations in neural compensation processes~\cite{gregory2017}. The results of the study incorporating graph theory analysis to resting-state fMRI data of prodromal Alzheimer’s disease patients also seem to confirm the involvement of the right hemisphere in compensation processes in the early stages of neurodegenerative diseases~\cite{behfar2020}. Despite the extensive investigation into neurodegeneration, a limited number of studies provide evidence of compensation, especially in the context of resting-state data. The results of our study suggest that we can detect brain cortical reorganization related to compensatory processes in the group of well-functioning RRMS patients using resting-state EEG data and multifractal analyses.

One of the possible implications of our research is the development of a methodology for the early diagnosis of patients with MS. In this light, the two aspects seem to be especially important. Firstly, the patients selected in our study had low levels of disability according to the EDSS scale, which, in our case, ranges between 0 and 3.5. This means that changes in the brain structure are not severe and they are challenging to detect. Our results demonstrate that even such early stages of the disease can be detected by multifractal analysis of EEG signals. It is worth emphasizing that, in general, the research in this area involves patients on a broad EDSS spectrum ~\cite{BABILONI2016581,vazquez2020} and/or with other MS phenotypes~\cite{kiiski2018,jamoussi2023}. Secondly, the presented study, refers to the EEG recording collected during resting state. Since the patients do not perform any tests during the recording, some cognitive deficits could be difficult to observe. It was demonstrated that multifractal analysis could quantify subtle changes in the temporal signal organisation, distinguishing the control and patient groups~\cite{keune2017exploring}.

\section*{Conclusions}
We conducted a correlation analysis, employing multifractal methodology, Pearson and detrended cross-correlation coefficients of electroencephalographic (EEG) data obtained from a multiple sclerosis (MS) study. The analysis concentrated on the relationship between the correlation characteristics, disease duration, and disability level. Its results indicate a correspondence between the complexity of EEG time series and the level of disability of MS patients quantified by the Expanded Disability Status Scale (EDSS). In particular, signals from MS patients with higher neurological impairment reveal multifractal organization, whereas monofractality characterizes control groups and patients with the lowest EDSS. In contrast, the persistence (linear correlations) is more pronounced for the combined group of control and patients with minimal signs of MS than for patients with observed disability. We hypothesize that the observed increase in the complexity of the EEG signals for MS patients is related to the brain’s compensatory processes and reflects structural brain complexity.

\section*{Acknowledgements}
This research was funded by the Foundation for Polish Science co-financed by the European Union under the European Regional Development Fund in the POIR.04.04.00-00-14DE/18-00 project carried out within the Team-Net programme. The research for this publication has been supported by a grant from the Priority Research Area DigiWorld under the Strategic Programme Excellence Initiative at Jagiellonian University.

\section*{Author contributions statement}
M.G., N.G-A., M.F., T.M., A.\.{Z}, M.M., M.W., and A.S., conceived and conducted the experiment and aided in data collection. M.G. and N.G-A. performed data preprocessing. M.W., J.O., W.T., and P.O. managed the statistical analysis, analysed the results, and prepared the figures. P.O., M.G., T.M. M.F, M.W., W.T., J.O., and N.G-A. interpreted the results and wrote and edited the manuscript. All authors reviewed the manuscript.

\section*{Data availability}
The EEG data analysed are available from Magda Gaw\l{}owska (email: magda.gawlowska@uj.edu.pl) on reasonable request.

\section*{Competing interests}
The authors declare no competing interests.

\makeatletter
\renewcommand*{\fnum@figure}{{\normalfont\bfseries \figurename~\thefigure}}
\makeatother

\makeatletter
\renewcommand*{\fnum@table}{{\normalfont\bfseries \tablename~\thetable}}
\makeatother
                
\renewcommand{\figurename}{Supplementary Fig.}
\renewcommand{\tablename}{Supplementary Table}

\renewcommand{\thefigure}{S\arabic{figure}}
\renewcommand{\thetable}{S\arabic{table}}
\renewcommand{\theequation}{S\arabic{equation}}

\newcounter{totfigures}
\newcounter{tottables}

\providecommand\totfig{} 
\providecommand\tottab{}

\makeatletter
\AtEndDocument{%
  \addtocounter{totfigures}{\value{figure}}%
  \addtocounter{tottables}{\value{table}}%
  \immediate\write\@mainaux{%
    \string\gdef\string\totfig{\number\value{totfigures}}%
    \string\gdef\string\tottab{\number\value{tottables}}%
  }%
}
\makeatother

\pretocmd{\chapter}{\addtocounter{totfigures}{\value{figure}}\setcounter{figure}{0}}{}{}
\pretocmd{\chapter}{\addtocounter{tottables}{\value{table}}\setcounter{table}{0}}{}{}

\section*{Supplementary Information}
\setcounter{figure}{0}
\begin{figure}[!hp]
\vspace{6cm}
  \includegraphics[width=.9\textwidth]{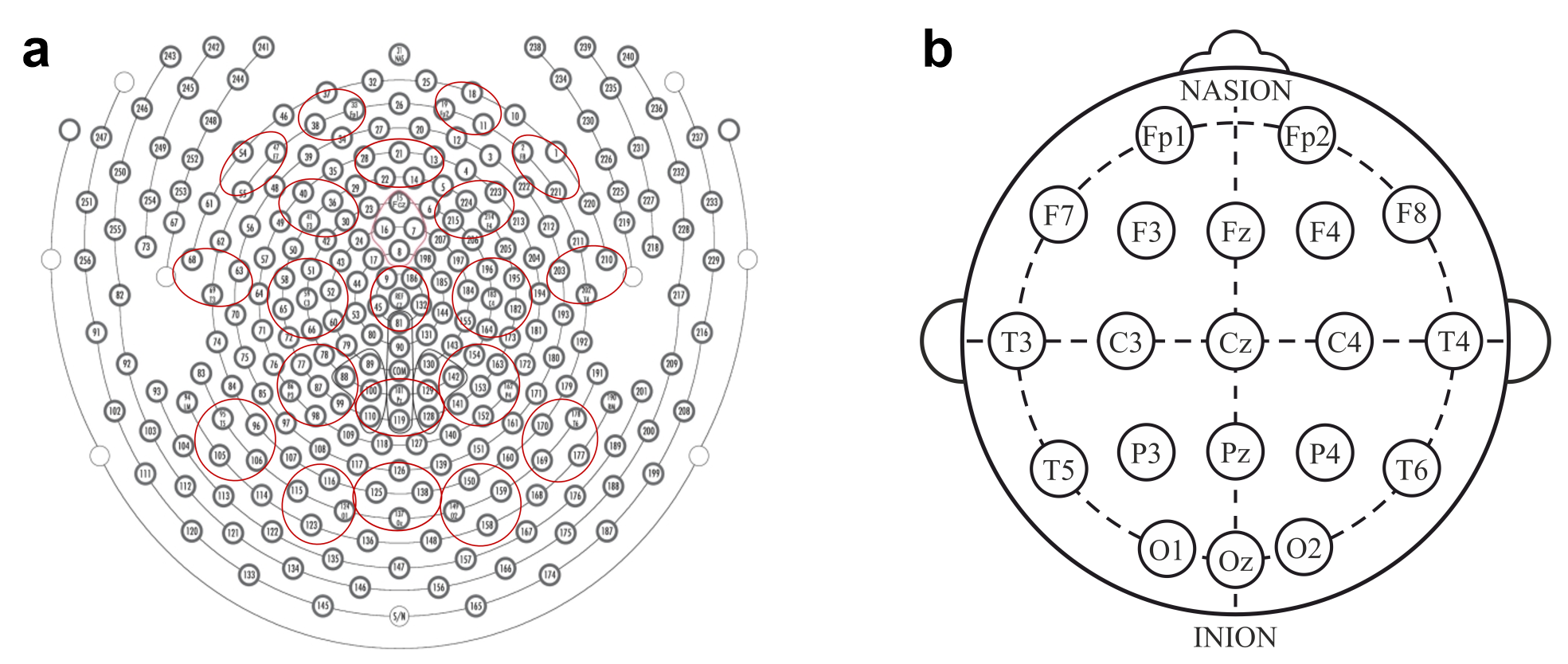}

\caption{\textbf{Diagrams of EEG electrodes analysed.} \textbf{a} Positions of 256 electrodes used in high-density recording with the HydroCel Geodesic Sensor Net, EGI System 300 (Electrical Geodesic Inc., OR, USA), with red circles grouping the electrodes into 10 -- 20 system. \textbf{b} The 10 -- 20 system electrodes used in the analyses. The electrode groups are also provided in Supplementary Table~\ref{tab::S_electrodemaps}.}
\label{fig::S1_electrodemaps}
\end{figure}

\newpage


\begin{figure}[htp!]
\centering
\includegraphics[width=0.65\textwidth]{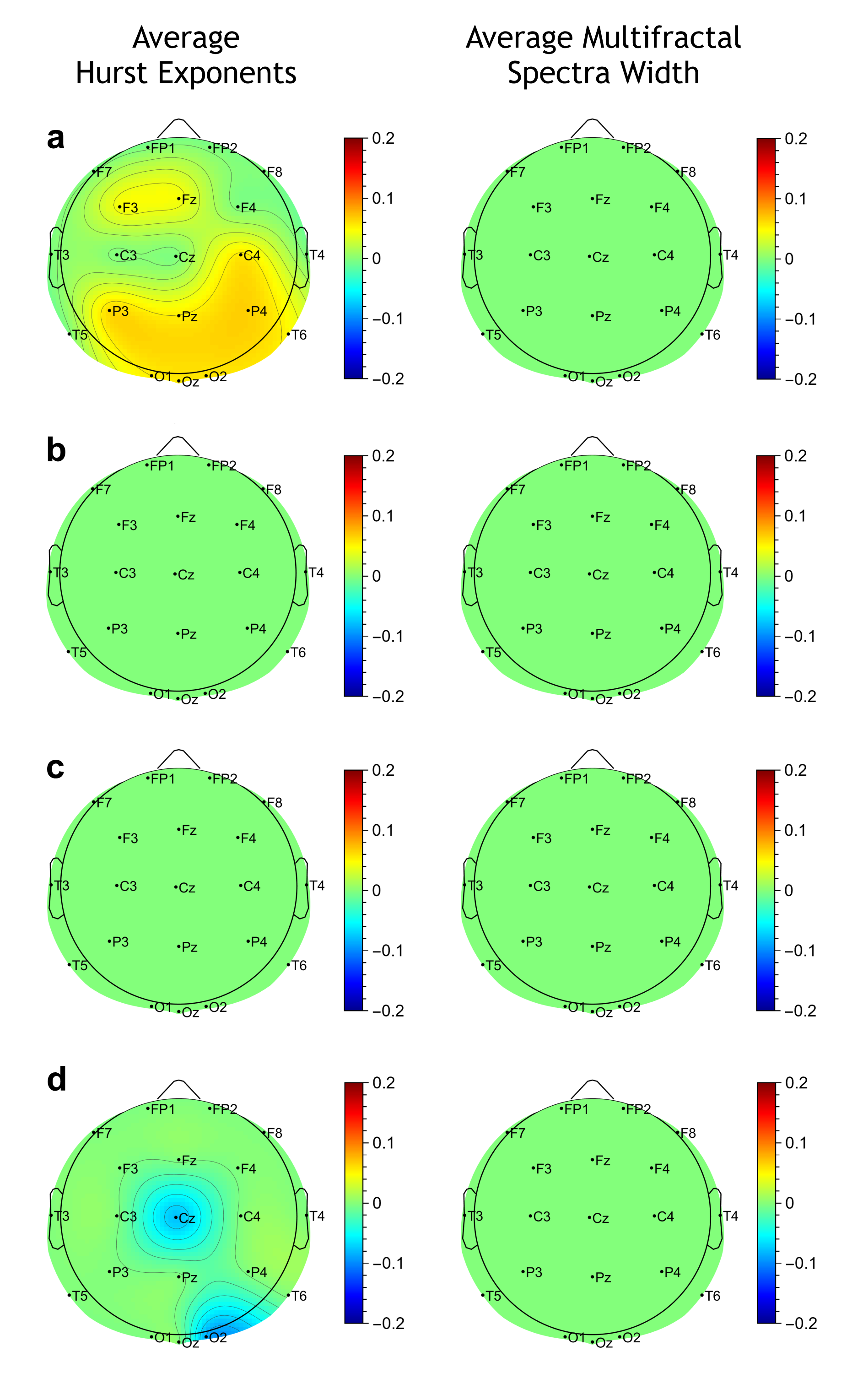}
\caption{\textbf{Group differences in Hurst exponents and multifractal spectrum widths -- eyes open.} The~topographic plots show statistically significant differences between: (\textbf{Left column}) group-averaged Hurst exponents $H$, and (\textbf{Right column}) multifractal spectra width $\Delta\alpha$.
The group comparisons include:
\textbf{a} the control group and patients,
\textbf{b} patients with $\textrm{EDSS}>1$ and patients with $\textrm{EDSS}\le1$,
\textbf{c} patients with disease duration longer than 7.5 years and shorter than 7.5 years.
\textbf{d} patients with $\textrm{EDSS}>1$ and the combined group of patients with $\textrm{EDSS}\le1$ and controls. These plots complement Figs.~4-5 from the main text with the open eyes condition.}
\label{fig::S_hurstmaps}
\end{figure}

   \begin{figure}[htp!]
\centering
\includegraphics[width=.89\textwidth]{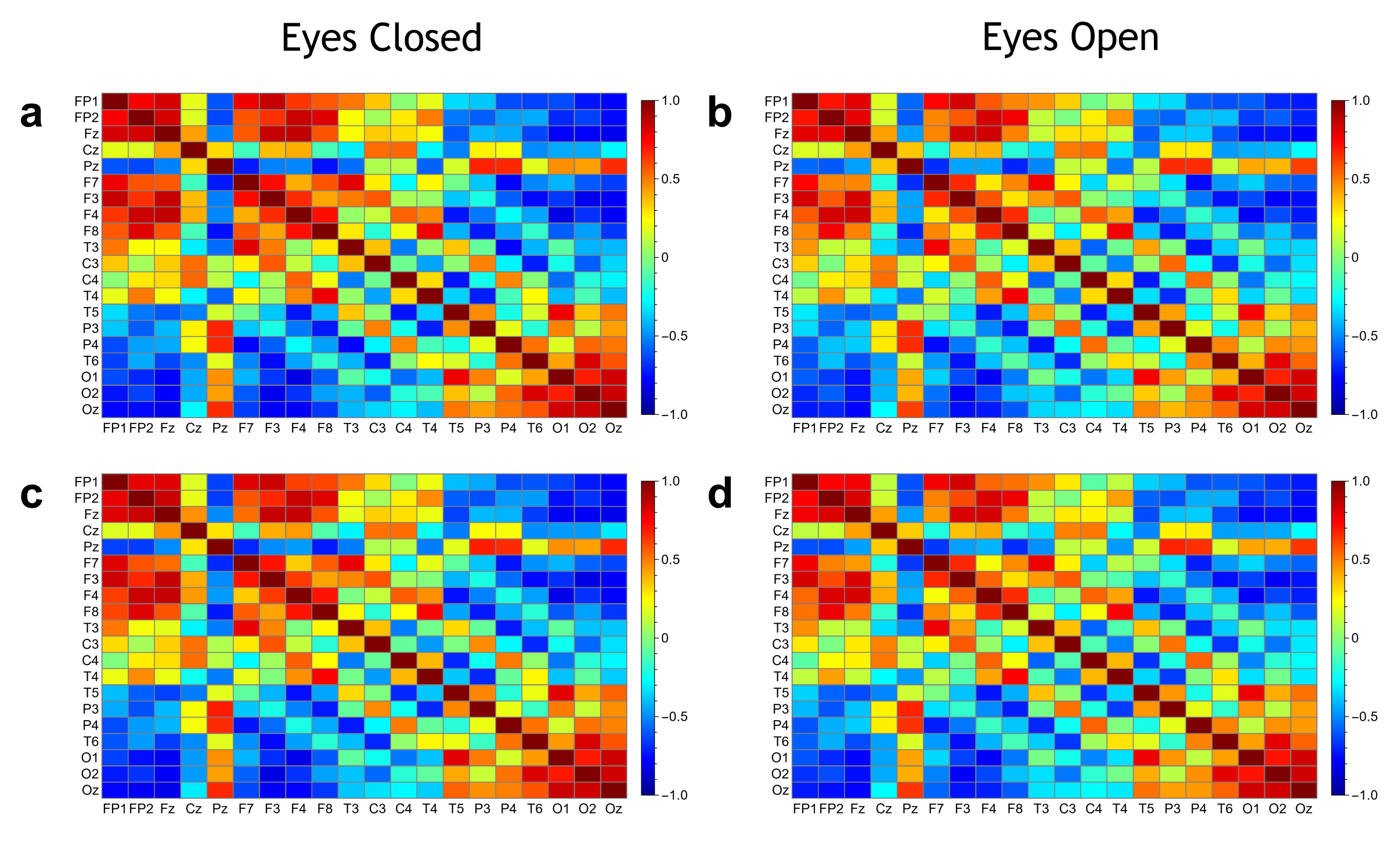}
\caption{\textbf{Electrode cross-correlations -- Pearson estimator.} These plots complement Fig.~6 from the main text with group-averaged Pearson correlation matrices for eyes closed and eyes open. \textbf{a-b} Control group. \textbf{c-d} Patients.}
\label{fig::S_Pearson-corr-mat}
\end{figure}
       \begin{figure}[htp!]
\centering
\includegraphics[width=.89\textwidth]{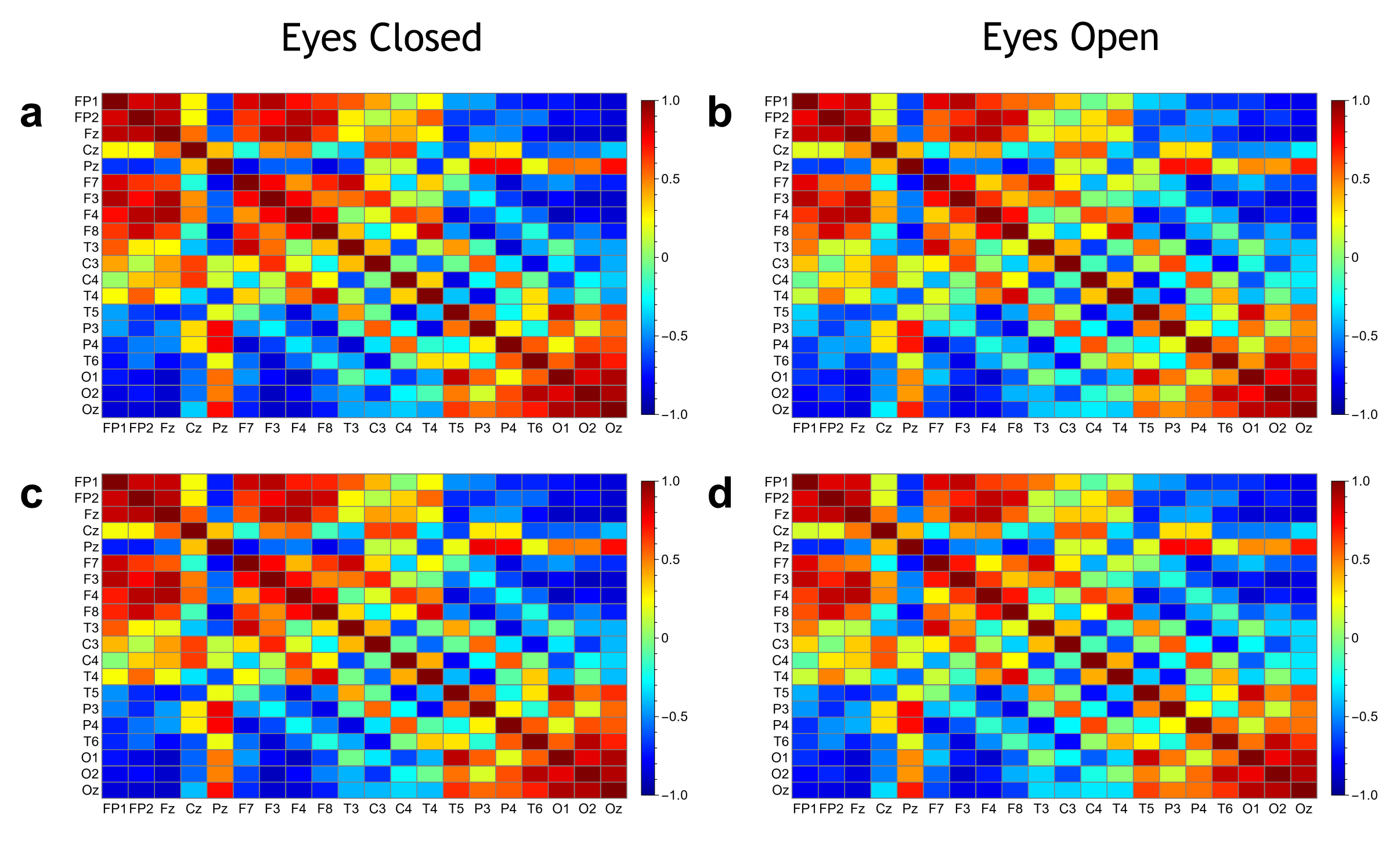}
\caption{\textbf{Electrode cross-correlations -- Detrended Cross-Correlation Coefficient.} These plots complement Fig.~6 from the main text (with the scale $s=200\,\textrm{ms}$) with group-averaged correlation matrices of $\rho(q=1,s=400\,\textrm{ms})$. \textbf{a-b} Control group. \textbf{c-d} Patients.}
\label{fig::S_Rho-corr-mat}
\end{figure}

\begin{figure}[htp!]
\includegraphics[width=0.735\textwidth]{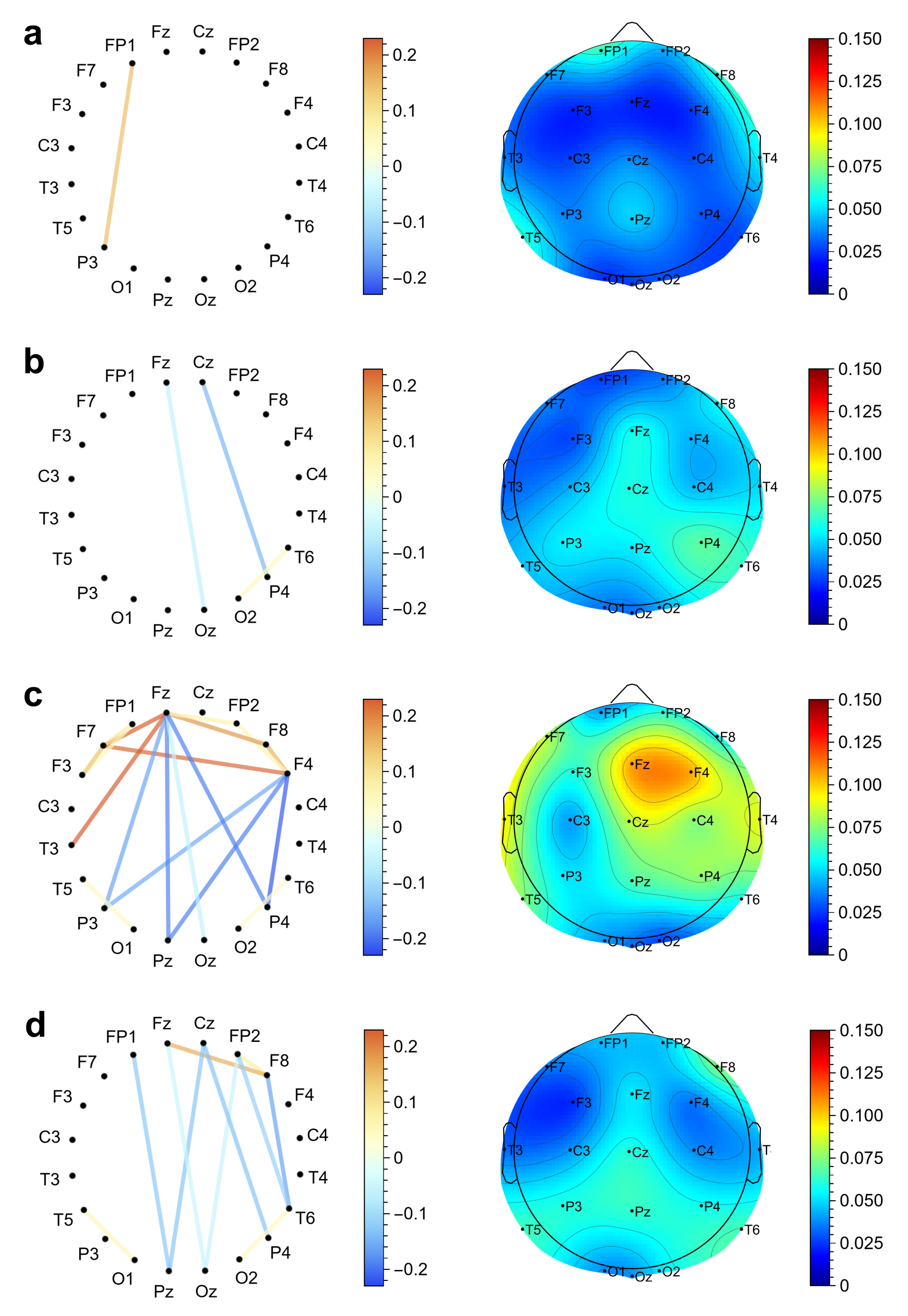}
\caption{\textbf{Group differences in detrended cross-correlations -- eyes open.}
(\textbf{Left column}) The links in the graph represent statistically significant group differences in detrended cross-correlations $\rho(q=1,s=200\,\textrm{ms})$ between pairs of electrodes (see Supplementary Table~\ref{tab::q_vals} for their estimated false discovery rates). Link colours indicate the value of the difference.
(\textbf{Right column}) The topographic plots show the absolute differences at each electrode averaged over cross-correlations with all other electrodes. The group comparisons include:
\textbf{a} the control group and patients,
\textbf{b} patients with $\textrm{EDSS}>1$ and patients with $\textrm{EDSS}\le1$,
\textbf{c} patients with the disease duration $\geq$ 7.5 and $<7.5$ years, 
\textbf{d} patients with $\textrm{EDSS}>1$ and the combined group of patients with $\textrm{EDSS}\le1$ and controls.
These plots complement Fig.~7 from the main text with the open eyes condition.}
\label{fig::S_networks-eyes-open}
\end{figure}
\newpage
\begin{figure}[htp!]
\centering
\includegraphics[width=0.93\textwidth]{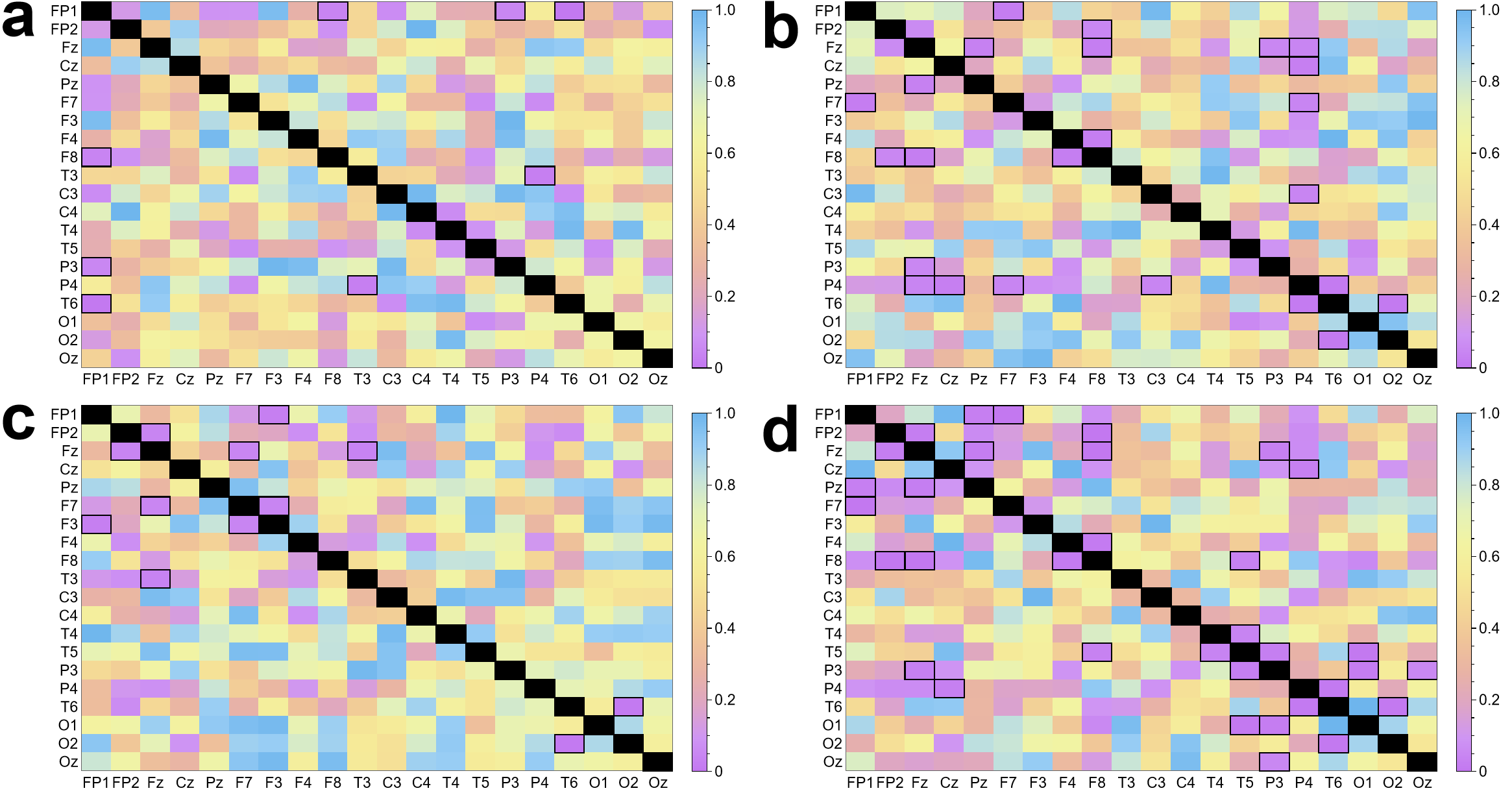}
\caption{\textbf{$p$-values for group differences in detrended cross-correlations -- eyes open.}
Each color-coded matrix corresponds to a group comparison from Supplementary Fig.~\ref{fig::S_networks-eyes-open}.
Each matrix element represents a $p$-value of Welch's $t$-test for a difference in detrended cross-correlations $\rho(q=1,s=200\,\textrm{ms})$ between a pair of electrodes.
Black borders indicate values $p < 0.05$, which are visible as links in the left-column graphs in Supplementary Fig.~\ref{fig::S_networks-eyes-open}. See Supplementary Table~\ref{tab::q_vals} for the estimated false discovery rates of these tests.
}
\label{fig::S_networks-open-pvals}
\end{figure}

\begin{figure}[htp!]
\centering
\includegraphics[width=0.93\textwidth]{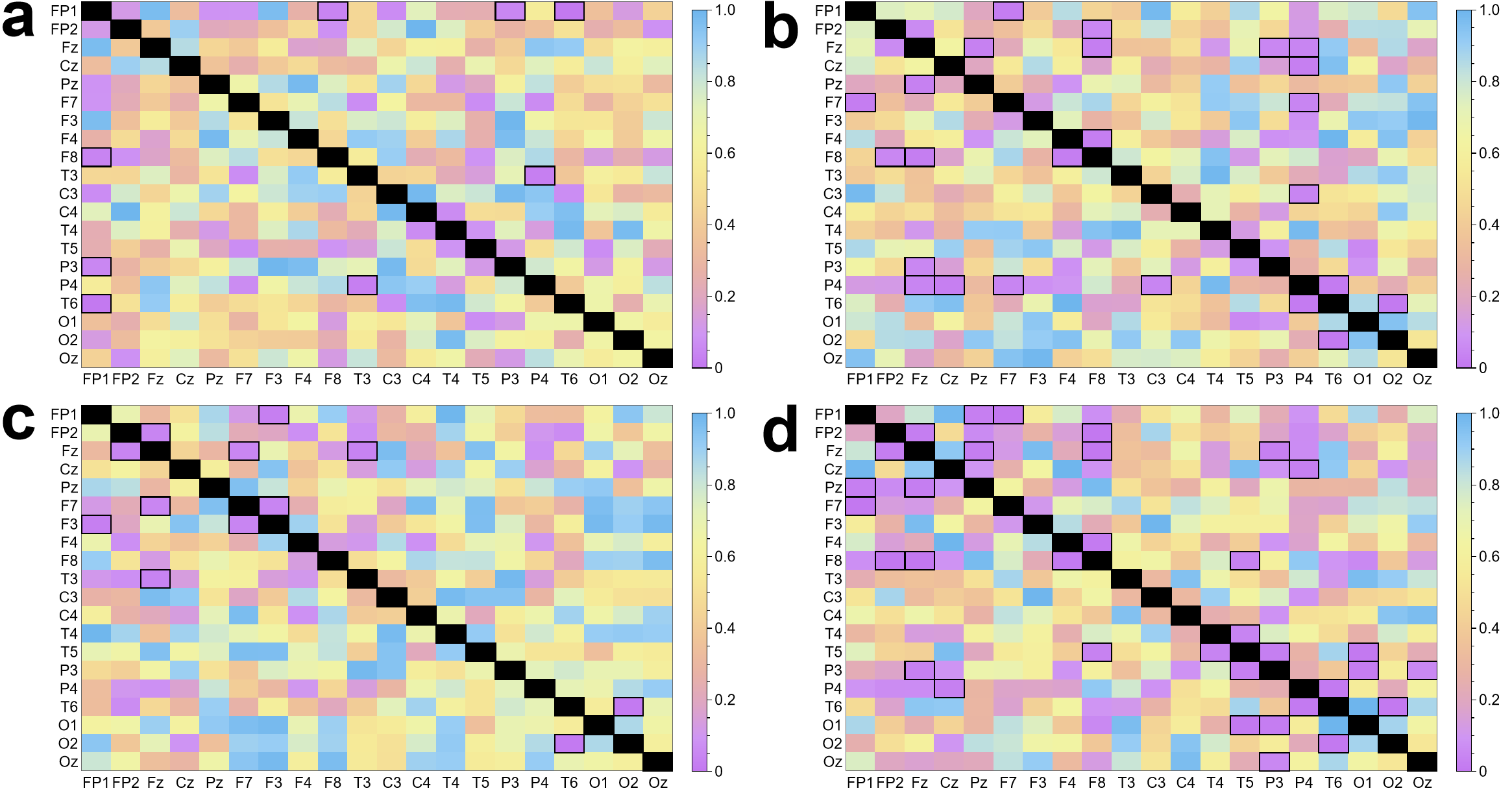}
\caption{\textbf{$p$-values for group differences in detrended cross-correlations -- eyes closed.}
Each color-coded matrix corresponds to a group comparison from main Fig.~7.
Each matrix element represents a $p$-value of Welch's $t$-test for a difference in detrended cross-correlations $\rho(q=1,s=200\,\textrm{ms})$ between a pair of electrodes.
Black borders indicate values $p < 0.05$, which are visible as links in the left-column graphs in main Fig.~7.
See Supplementary Table~\ref{tab::q_vals} for the estimated false discovery rates of these tests.
}
\label{fig::S_networks-close-pvals}
\end{figure}

\begin{figure}[htp!]
\centering
\includegraphics[width=0.75\textwidth]{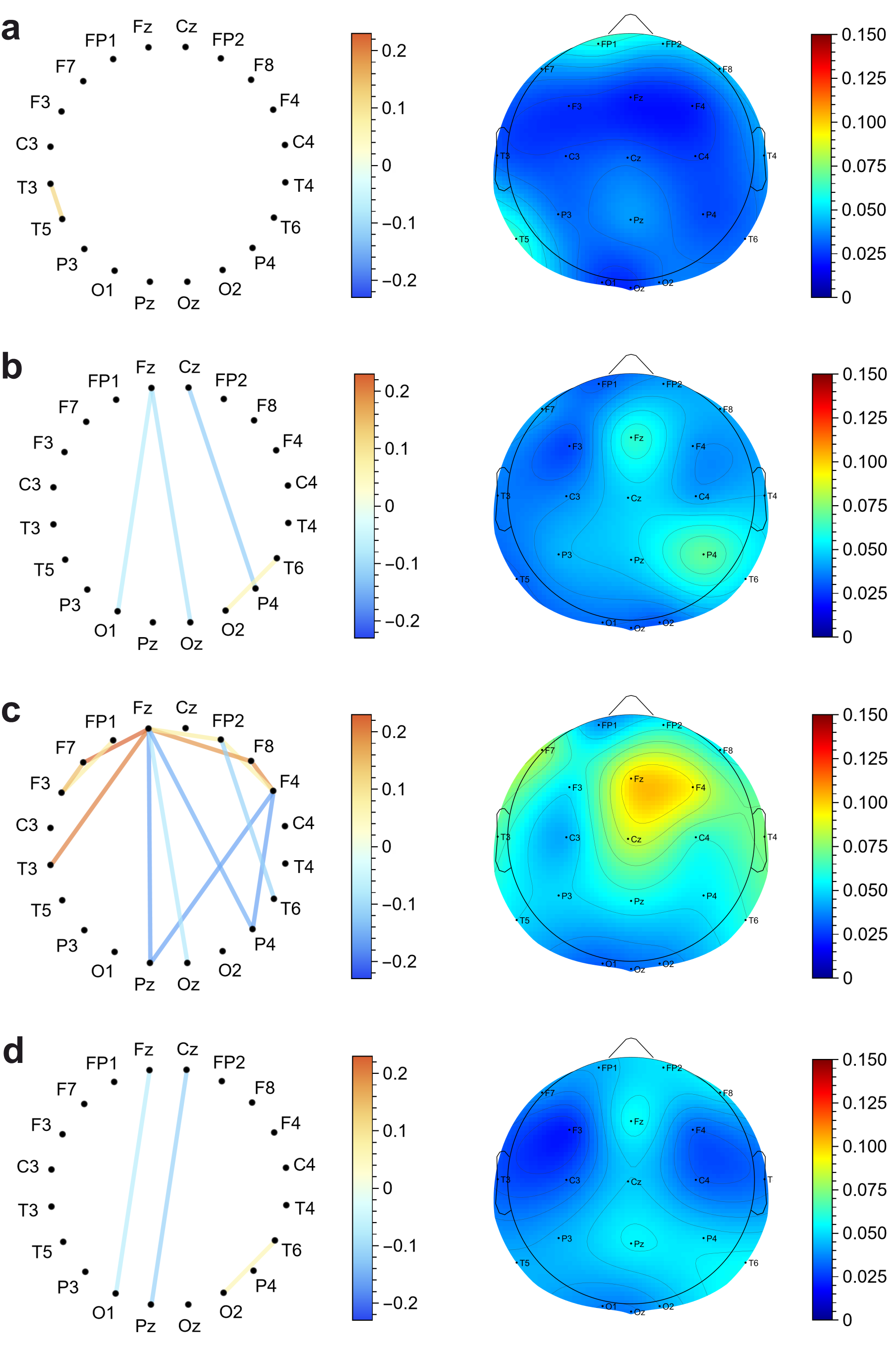}
\caption{\textbf{Group differences in Pearson cross-correlations -- eyes open.}
(\textbf{Left column}) The links in the graph represent statistically significant group differences in Pearson correlations between pairs of electrodes. Link colours indicate the value of the difference.
(\textbf{Right column}) The topographic plots show the absolute differences at each electrode averaged over cross-correlations with all other electrodes.
For comparison with detrended cross-correlations, see Supplementary Fig.~\ref{fig::S_networks-eyes-open}.}
\label{fig::S_networks-pearson-open}
\end{figure}

\begin{figure}[htp!]
\centering
\includegraphics[width=0.75\textwidth]{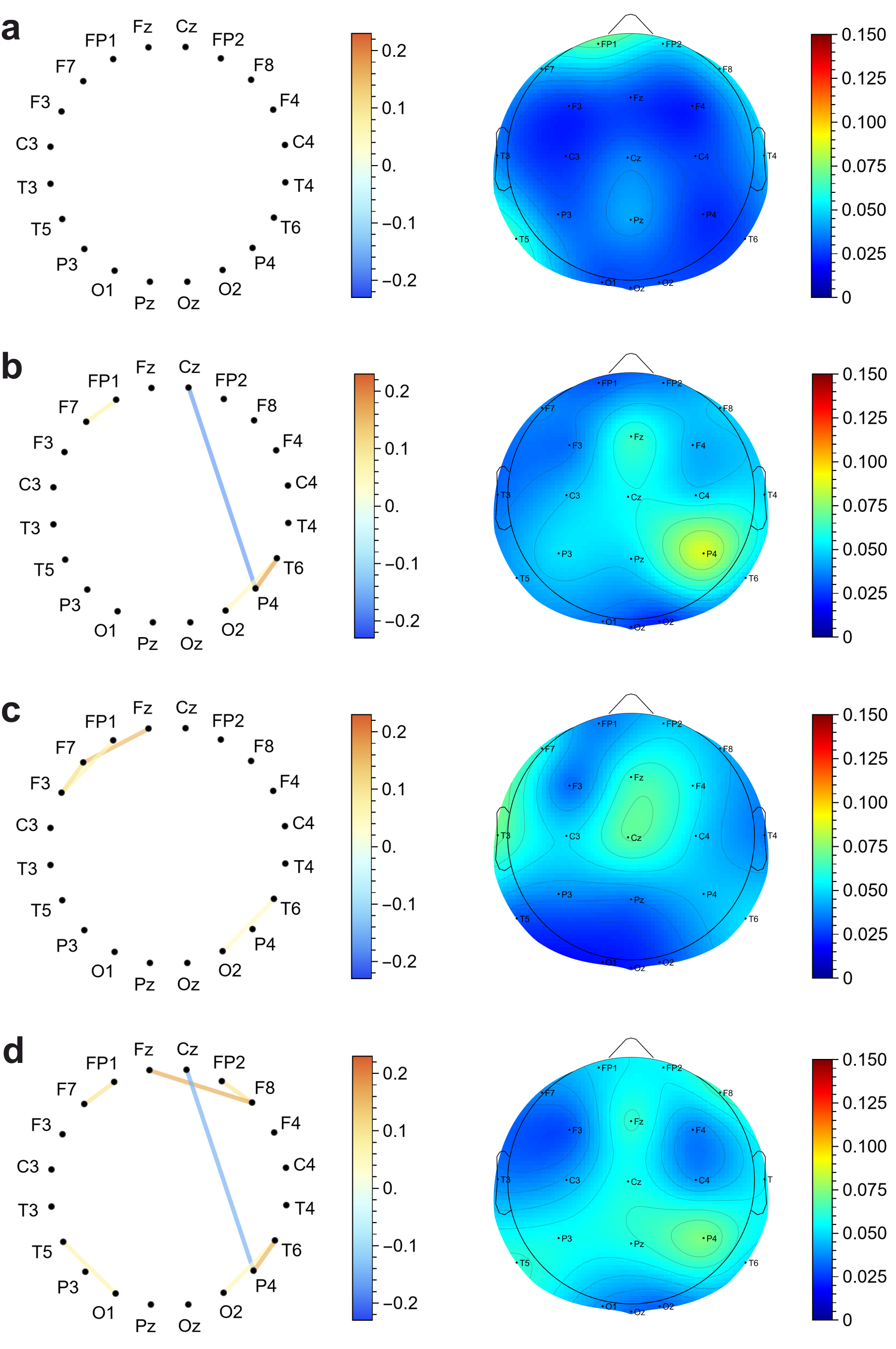}
\caption{\textbf{Group differences in Pearson cross-correlations -- eyes closed.}
(\textbf{Left column}) The links in the graph represent statistically significant group differences in Pearson correlations between pairs of electrodes. Link colours indicate the value of the difference.
(\textbf{Right column}) The topographic plots show the absolute differences at each electrode averaged over cross-correlations with all other electrodes.
For comparison with detrended cross-correlations, see main Fig.~7.}
\label{fig::S_networks-pearson-close}
\end{figure}
\newpage
\begin{figure}[htbp!]
\centering
\includegraphics[width=0.95\textwidth]{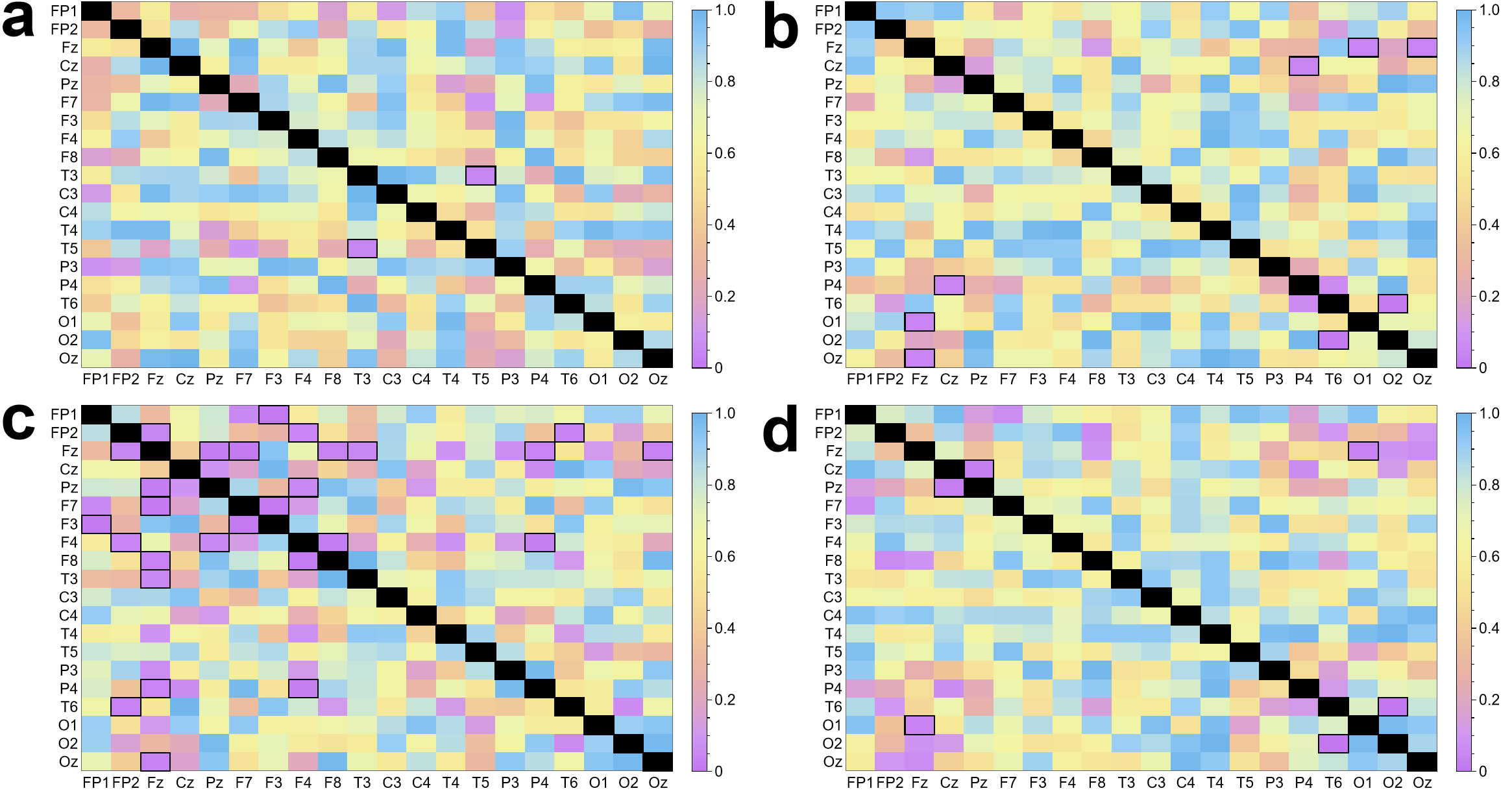}
\caption{\textbf{$p$-values for group differences in Pearson cross-correlations -- eyes open.}
Each color-coded matrix corresponds to a group comparison from Supplementary Fig.~\ref{fig::S_networks-pearson-open}.
Each matrix element represents a $p$-value of Welch's $t$-test for a difference in Pearson correlations between a pair of electrodes.
Black borders indicate values $p < 0.05$, which are visible as links in the left-column graphs in Supplementary Fig.~\ref{fig::S_networks-pearson-open}.
}
\label{fig::S_pearson-open-pvals}
\end{figure}
\begin{figure}[htbp!]
\centering
\includegraphics[width=0.95\textwidth]{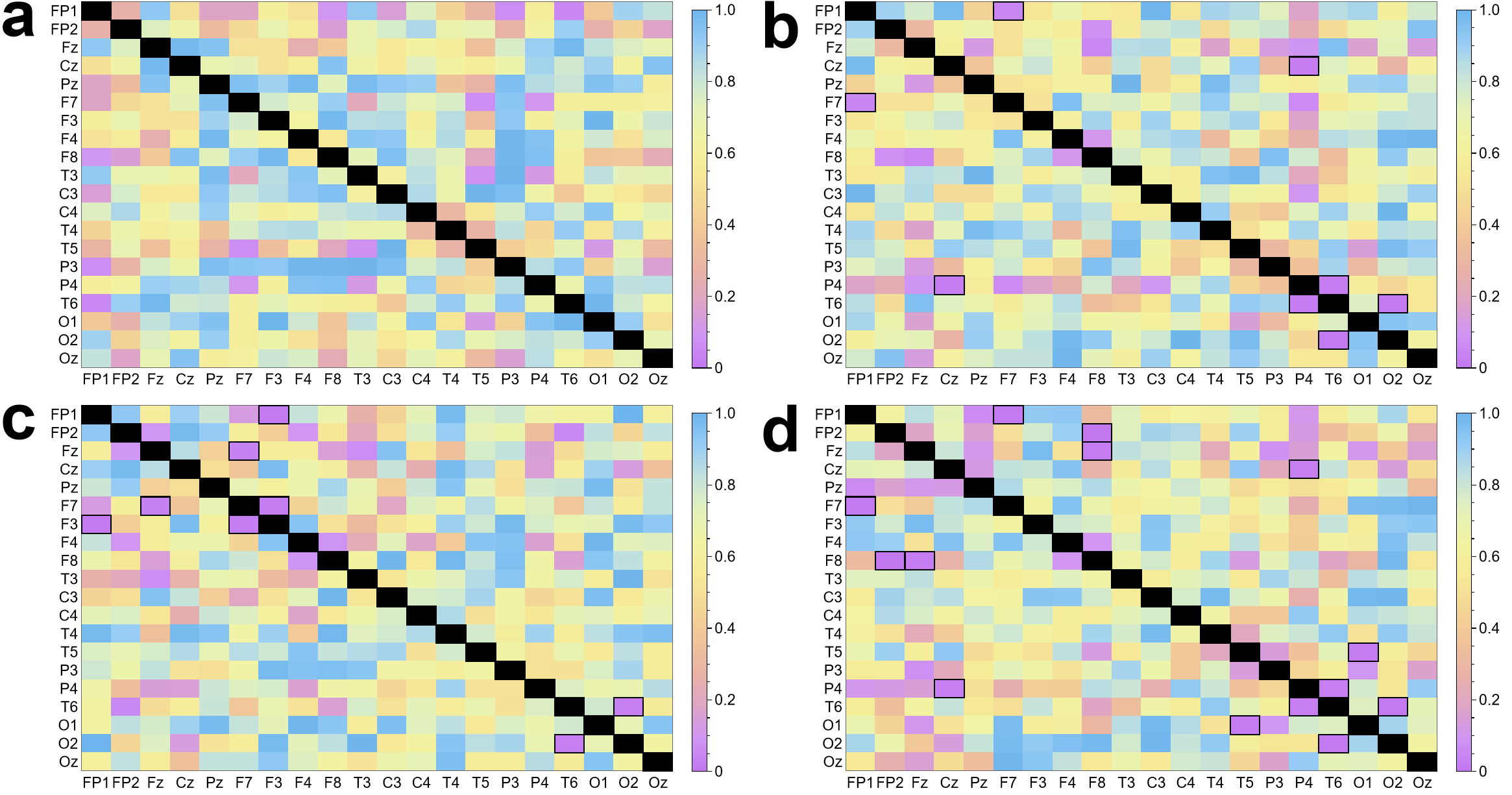}
\caption{\textbf{$p$-values for group differences in Pearson cross-correlations -- eyes closed.}
Each color-coded matrix corresponds to a group comparison from Supplementary Fig.~\ref{fig::S_networks-pearson-close}.
Each matrix element represents a $p$-value of Welch's $t$-test for a difference in Pearson correlations between a pair of electrodes.
Black borders indicate values $p < 0.05$, which are visible as links in the left-column graphs in Supplementary Fig.~\ref{fig::S_networks-pearson-close}.
}
\label{fig::S_pearson-close-pvals}
\end{figure}
\newpage
\begin{table}[!ht]
    \centering
    \renewcommand{\arraystretch}{1.15}
    \caption{\textbf{Electrode groups.} EGI system electrodes groups corresponding to each of the 10 -- 20 system electrodes as presented in Supplementary Fig.~\ref{fig::S1_electrodemaps}.}
    \vspace{0.5cm}
    \small
    \setlength{\tabcolsep}{2pt}
    \renewcommand{\arraystretch}{0.9}
    \begin{tabular}{r@{\hskip5pt}|lllllll}
    
        \textbf{10 -- 20} & \multicolumn{7}{c}{\textbf{EGI system}} \\ \hline
        \textbf{Fz} & 13 & 14 & 21 & 22 & 28 & ~ & ~ \\ 
        \textbf{Cz} & 9 & 186 & 45 & 132 & 81 & ~ & ~ \\ 
        \textbf{Pz} & 101 & 100 & 129 & 110 & 119 & 128 & ~ \\ 
        \textbf{Oz} & 126 & 125 & 138 & 137 & ~ & ~ & ~ \\
        \textbf{FP1} & 37 & 38 & 33 & ~ & ~ & ~ & ~ \\ 
        \textbf{FP2} & 18 & 19 & 11 & ~ & ~ & ~ & ~ \\ 
        \textbf{F3} & 36 & 30 & 40 & 41 & ~ & ~ & ~ \\ 
        \textbf{F7} & 47 & 54 & 55 & ~ & ~ & ~ & ~ \\ 
        \textbf{F4} & 224 & 215 & 223 & 214 & ~ & ~ & ~ \\ 
        \textbf{F8} & 2 & 1 & 221 & ~ & ~ & ~ & ~ \\ 
        \textbf{T3} & 69 & 68 & 63 & ~ & ~ & ~ & ~ \\ 
        \textbf{T5} & 96 & 95 & 105 & 106 & ~ & ~ & ~ \\ 
        \textbf{T4} & 202 & 203 & 210 & ~ & ~ & ~ & ~ \\ 
        \textbf{T6} & 170 & 169 & 177 & 178 & ~ & ~ & ~ \\ 
        \textbf{C3} & 59 & 51 & 52 & 60 & 66 & 65 & 58 \\ 
        \textbf{C4} & 183 & 184 & 182 & 155 & 164 & 196 & 195 \\ 
        \textbf{P3} & 87 & 77 & 78 & 88 & 86 & 98 & 99 \\ 
        \textbf{P4} & 153 & 141 & 142 & 152 & 154 & 162 & 163 \\ 
        \textbf{O1} & 116 & 115 & 123 & 124 & ~ & ~ & ~ \\ 
        \textbf{O2} & 150 & 149 & 158 & 159 & ~ & ~ & ~ \\
    \end{tabular}

\label{tab::S_electrodemaps}
\end{table}
\newpage
\begin{table}
\centering
\caption{\textbf{Results of $t$-tests.} The table presents $p$-values for the results presented in Fig.~4-5 in the main text. The group comparisons include:
\textbf{a} the control group and patients,
\textbf{b} patients with $\textrm{EDSS}>1$ and patients with $\textrm{EDSS}\le1$,
\textbf{c} patients with the disease duration $\geq$ 7.5 and $<7.5$ years, 
\textbf{d} patients with $\textrm{EDSS}>1$ and the combined group of patients with $\textrm{EDSS}\le1$ and controls.
The results are shown for the closed eyes condition. $p$-values $\leq 0.05$ are in bold. MF stands for ``multifractal''.}
    \vspace{0.5cm}
\small

\setlength{\tabcolsep}{2pt}
\begin{tabular}{l|l|llllllllllllllllllll}
\multicolumn{2}{l|}{}       & FP1           & FP2           & Fz            & Cz            & Pz            & F7            & F3            & F4            & F8            & T3            & C3            & C4            & T4            & T5   & P3            & P4            & T6            & O1            & O2            & Oz             \\ 
\hline
\multirow{4}{*}{\rotatebox[origin=c]{90}{\shortstack{Hurst\\Exponents}}}  & \textbf{a} & 0.26          & 0.26          & 0.21          & 0.13          & \textbf{0.04} & 0.14          & 0.08          & 0.22          & 0.13          & 0.10          & 0.06          & 0.20          & \textbf{0.04} & 0.30 & 0.07          & 0.07          & 0.15          & 0.18          & 0.08          & 0.08           \\
                        & \textbf{b} & \textbf{0.03} & \textbf{0.05} & 0.05          & \textbf{0.02} & 0.15          & \textbf{0.01} & \textbf{0.03} & 0.07          & \textbf{0.05} & 0.07          & 0.11          & \textbf{0.05} & 0.09          & 0.13 & 0.19          & 0.05          & 0.10          & 0.10          & 0.07          & 0.08           \\
                        & \textbf{c} & 0.49          & 0.49          & 0.70          & 0.96          & 0.75          & 0.57          & 0.93          & 0.54          & 0.74          & 0.52          & 0.54          & 0.69          & 0.72          & 0.93 & 0.79          & 0.99          & 0.72          & 0.68          & 0.76          & 0.45           \\
                        & \textbf{d} & \textbf{0.02} & \textbf{0.03} & \textbf{0.03} & \textbf{0.01} & \textbf{0.04} & \textbf{0.01} & \textbf{0.01} & \textbf{0.04} & \textbf{0.02} & \textbf{0.03} & \textbf{0.04} & \textbf{0.02} & \textbf{0.03} & 0.05 & 0.06          & \textbf{0.01} & \textbf{0.03} & \textbf{0.04} & \textbf{0.02} & \textbf{0.03}  \\ 
\hline
\multirow{4}{*}{\rotatebox[origin=c]{90}{\shortstack{MF Spectra\\  Widths}
}} & \textbf{a} & 0.78          & 0.88          & 0.88          & 0.87          & 0.33          & 0.80          & 0.72          & 0.53          & 0.52          & 0.28          & 0.34          & 0.45          & 0.40          & 0.90 & \textbf{0.04} & 0.64          & 0.26          & 0.42          & 0.19          & 0.36           \\
                        & \textbf{b} & \textbf{0.01} & \textbf{0.03} & 0.09          & \textbf{0.05} & 0.14          & \textbf{0.04} & \textbf{0.04} & 0.14          & 0.15          & \textbf{0.01} & 0.06          & 0.15          & 0.10          & 0.55 & 0.26          & \textbf{0.03} & \textbf{0.04} & 0.45          & 0.09          & 0.09           \\
                        & \textbf{c} & 0.06          & \textbf{0.01} & 0.10          & 0.58          & 0.17          & 0.24          & 0.20          & 0.07          & 0.17          & 0.25          & 0.62          & 0.33          & 0.91          & 0.44 & 0.82          & 0.52          & 0.94          & 0.13          & 0.74          & 0.13           \\
                        & \textbf{d} & \textbf{0.03} & 0.05          & 0.13          & 0.06          & 0.11          & 0.07          & 0.05          & 0.24          & 0.16          & 0.08          & \textbf{0.04} & 0.31          & 0.23          & 0.64 & 0.18          & \textbf{0.04} & \textbf{0.02} & 0.39          & 0.07          & 0.08          
\end{tabular}

\label{tab::S_p=vals}
\end{table}

\newpage
\begin{table}
\caption{\textbf{Results of $t$-tests.} The table presents $p$-values for the results presented in Supplementary Fig.~\ref{fig::S_hurstmaps}. The group comparisons include:
\textbf{a} the control group and patients,
\textbf{b} patients with $\textrm{EDSS}>1$ and patients with $\textrm{EDSS}\le1$,
\textbf{c} patients with the disease duration $\geq$ 7.5 and $<7.5$ years, 
\textbf{d} patients with $\textrm{EDSS}>1$ and the combined group of patients with $\textrm{EDSS}\le1$ and controls.
The results are shown for the open eyes condition. $p$-values $\leq 0.05$ are in bold. MF stands for ``multifractal''.}
    \vspace{0.5cm}
\small
\setlength{\tabcolsep}{2pt}
\begin{tabular}{ll|llllllllllllllllllll}
\multicolumn{2}{l|}{\textbf{}}                                                                                             & FP1  & FP2  & Fz            & Cz            & Pz            & F7   & F3            & F4   & F8   & T3   & C3   & C4            & T4   & T5   & P3            & P4            & T6            & O1            & O2            & Oz            \\ \hline
\multicolumn{1}{l|}{\multirow{4}{*}{\rotatebox[origin=c]{90}{\shortstack{Hurst\\ Exponents}}}}              & \textbf{a} & 0.09 & 0.09 & \textbf{0.04} & 0.07          & \textbf{0.03} & 0.11 & \textbf{0.04} & 0.06 & 0.10 & 0.21 & 0.08 & \textbf{0.03} & 0.10 & 0.06 & \textbf{0.03} & \textbf{0.02} & \textbf{0.04} & \textbf{0.01} & \textbf{0.01} & \textbf{0.01} \\
\multicolumn{1}{l|}{}                                                                                         & \textbf{b} & 0.24 & 0.32 & 0.35          & 0.14          & 0.55          & 0.14 & 0.20          & 0.38 & 0.30 & 0.39 & 0.18 & 0.38          & 0.26 & 0.46 & 0.32          & 0.31          & 0.23          & 0.28          & 0.16          & 0.24          \\
\multicolumn{1}{l|}{}                                                                                         & \textbf{c} & 0.98 & 0.90 & 0.72          & 0.65          & 0.74          & 0.77 & 0.77          & 0.80 & 0.72 & 0.90 & 0.41 & 0.54          & 0.38 & 0.42 & 0.50          & 0.66          & 0.44          & 0.74          & 0.77          & 0.86          \\
\multicolumn{1}{l|}{}                                                                                         & \textbf{d} & 0.12 & 0.18 & 0.14          & \textbf{0.04} & 0.20          & 0.07 & 0.08          & 0.18 & 0.12 & 0.24 & 0.09 & 0.15          & 0.14 & 0.18 & 0.11          & 0.10          & 0.07          & 0.09          & \textbf{0.04} & 0.07          \\ \hline
\multicolumn{1}{l|}{\multirow{4}{*}{\rotatebox[origin=c]{90}{\shortstack{MF Spectra\\ Widths\;}}}} & \textbf{a} & 0.25 & 0.09 & 0.54          & 0.46          & 0.83          & 0.70 & 0.97          & 0.49 & 0.87 & 0.66 & 0.98 & 0.90          & 0.91 & 0.56 & 0.47          & 0.50          & 0.65          & 0.92          & 0.70          & 0.95          \\
\multicolumn{1}{l|}{}                                                                                         & \textbf{b} & 0.70 & 0.33 & 0.81          & 0.87          & 0.87          & 0.73 & 0.17          & 0.33 & 0.95 & 0.78 & 0.32 & 0.27          & 0.50 & 0.06 & 0.93          & 0.63          & 0.87          & 0.29          & 0.17          & 0.48          \\
\multicolumn{1}{l|}{}                                                                                         & \textbf{c} & 0.44 & 0.23 & 0.56          & 0.15          & 0.30          & 0.23 & 0.63          & 0.98 & 0.82 & 0.05 & 0.07 & 0.55          & 0.83 & 0.98 & 0.21          & 0.39          & 0.12          & 0.91          & 0.73          & 0.95          \\
\multicolumn{1}{l|}{}                                                                                         & \textbf{d} & 0.72 & 0.79 & 0.96          & 0.84          & 0.79          & 0.66 & 0.22          & 0.59 & 0.94 & 0.91 & 0.33 & 0.35          & 0.60 & 0.23 & 0.74          & 0.83          & 0.69          & 0.34          & 0.17          & 0.55         
\end{tabular}
\label{tab::S_p=vals_open}
\end{table}

\begin{table}
\caption{\textbf{False discovery rate for detrended cross-correlations.} The table presents the maximum estimated $q$-value (estimated false discovery rate) among all $p$-values $\leq 0.05$ for the results presented in Fig.~8 and in Supplementary Fig.~\ref{fig::S_networks-eyes-open}-\ref{fig::S_networks-close-pvals}. The group comparisons include:
\textbf{a} the control group and patients,
\textbf{b} patients with $\textrm{EDSS}>1$ and patients with $\textrm{EDSS}\le1$,
\textbf{c} patients with the disease duration $\geq$ 7.5 and $<7.5$ years, 
\textbf{d} patients with $\textrm{EDSS}>1$ and the combined group of patients with $\textrm{EDSS}\le1$ and controls.}
    \vspace{0.5cm}
    \small
\begin{tabular}{l|llll|llll}
       & \multicolumn{4}{c|}{s = 200 ms}                        & \multicolumn{4}{c}{s = 400 ms}                        \\ \cline{2-9} 
       & \textbf{a} & \textbf{b} & \textbf{c} & \textbf{d} & \textbf{a} & \textbf{b} & \textbf{c} & \textbf{d} \\ \hline
Eyes Closed   & 0.94       & 0.76       & 0.99       & 0.53       & 0.96       & 0.88       & 1.         & 0.66       \\
Eyes Open & 1.         & 0.99       & 0.34       & 0.76       & 1.         & 0.94       & 0.21       & 0.81      
\end{tabular}
\label{tab::q_vals}
\end{table}

\bibliography{References} 
\bibliographystyle{elsarticle-num-names} 

\end{document}